\documentclass[
aps,%
11pt,%
final,%
notitlepage,%
oneside,%
onecolumn,%
nobibnotes,%
nofootinbib,%
superscriptaddress,%
noshowpacs,%
centertags]%
{revtex4}

\begin{document}
\selectlanguage{english}

\title{Analysis of the Properties of Clusters of Galaxies in the
Region \protect \linebreak of the Ursa Major Supercluster}
\author{\firstname{F.~G.}~\surname{Kopylova}}
\affiliation{\saoname}
\author{\firstname{A.~I.}~\surname{Kopylov}}
\affiliation{\saoname}

\received{August 20, 2008}%
\revised{September 25, 2008}%

\begin{abstract}
We analyze the properties of the clusters of galaxies in the region of the Ursa
Major (UMa) supercluster using observational data from SDSS and 2MASS catalogs.
The region studied includes a supercluster (with a galaxy and cluster
overdensity of 3 and 15, respectively) and field clusters inside the
150-Mpc diameter surrounding region. The total
dynamical mass of 10
clusters of galaxies in UMa is equal to $2.25 \times 10^{15}$ $M_{\odot}$,
and the mass of 11 clusters of galaxies in the UMa neighborhood is equal to
\mbox{$1.70 \times 10^{15}$ $M_{\odot}$.} The fraction of early-type galaxies brighter
than $M_K^*+1$ in the virialized regions of clusters is, on the average, equal
to $70\%$, and it is virtually independent on the mass of the cluster. The
fraction of these galaxies and their average photometric parameters are almost
the same both for UMa clusters and for the clusters located in its surroundings.
Parameters of the clusters of galaxies, such as infrared luminosities up to a fixed
magnitude, the mass-to-luminosity ratio, and the number of galaxies have almost
the same correlations with the cluster mass as in other samples of galaxies clusters.
However, the scatter of these parameters for UMa member clusters is
twice smaller than the corresponding scatter
for field clusters, possibly, due to the common origin of UMa clusters and
synchronized dynamical evolution of clusters in the supercluster.

\end{abstract}

\maketitle

\section{INTRODUCTION}

Clusters of galaxies, which are the biggest virialized systems in the Universe,
form even larger structures---the superclusters. Superclusters consist of two
to twenty clusters and groups of galaxies, which are located either in filaments
or at the intersections of filaments (nodes). Superclusters are large structures
and their study requires the use of extensive observational data---first and
foremost, radial velocities of galaxies. For example, based on their analysis
of the 2dF survey, Einasto et al.~\cite{Einasto1:Kopylova_n} revealed the
following property of the superclusters of galaxies: rich clusters of
galaxies, located in the densest regions of rich superclusters, contain
greater fractions of early-type galaxies. Note that the brightest galaxies of
the main clusters (i.e., bright clusters located near the peak of
luminosity density) in rich superclusters are brighter than the corresponding
galaxies in poor superclusters or field groups. An analysis of the rich
supercluster Corona Borealis~\mbox{($z\sim0.07$,~\cite{Small:Kopylova_n})}
shows that this system exhibits an excess of bright galaxies compared
to the field. A study of the properties of galaxies in the Shapley supercluster
yielded the following results  \cite{Haines:Kopylova_n}: the
($B-R,R$) color-magnitude relation shows that early-type galaxies in the cores
of clusters are  $0\fm015$ redder (older) than in less dense regions, whereas
the fraction of late-type galaxies increases with decreasing local galaxy
density and with increasing magnitude. Moreover, the Schechter function fits
poorly the composite luminosity function of the supercluster
\cite{Mercurio:Kopylova_n}.

The aim of this paper is to compare the properties of clusters of galaxies and
the average parameters of the subsamples located within the virial radius: (1)
12 clusters located in the Ursa Major supercluster and (2) 12 isolated field
clusters located in the nearest neighborhood of the Ursa Major supercluster
within 75~Mpc of the center of the system. For this study we used the data of
the 2MASS (Two-Micron All-Sky Survey) and SDSS (Sloan Digital Sky Survey)
catalogs. The paper has the following layout. Part~2 describes the parameters of
clusters of galaxies: the dispersion of galaxy velocities, radius of the virial
region, and mass. In Part~3 we compute and list the total $K_s$-band (hereafter
simply referred to as $K$-band) luminosities of the clusters, the composite
luminosity function of the clusters of the system, the luminosity functions of
early- and late-type galaxies, and various correlations between the parameters
of clusters of galaxies. In Conclusions we list the results obtained. Throughout
this paper we adopt the following cosmological parameters:
$\Omega_m=0.3$, $\Omega_{\Lambda}=0.7$, and $H_0=70$~km/s/Mpc.

\section{PARAMETERS OF CLUSTERS}
The sample studied is made up of 12 member clusters of the UMa supercluster (the
supercluster is the space region where the galaxy overdensity is about~\mbox{3
\cite{Kop1:Kopylova_n}),} and 12 clusters of galaxies located in the UMa
neighborhood with a lower overdensity. The redshifts of clusters lie in the
interval \mbox{$0.045<z<0.075$.} We use for the galaxies studied the spectroscopic data
reported in the SDSS (Data Release 5)  catalog and supplement them by the data
adopted from NED. In our earlier papers we used the Data Release 3
\cite{Kop2:Kopylova_n} and Data Release 4 \cite{Kop1:Kopylova_n} catalogs, which
contained incomplete data for some of the clusters (A1291, A1377, A1436, and
Anon4). In this paper we refine all the parameters of the clusters of galaxies based on
SDSS data  (our previous papers were based on the 2MASS catalog).

\subsection{Dynamic Parameters of Clusters of Galaxies}
We determine the dynamical masses of  clusters from the dispersion of galaxy
radial velocities assuming that the cluster is in virial equilibrium and the masses
of galaxies increase linearly with radius.
According to Carlberg et al.~\cite{Carlberg:Kopylova_n}, the radius of a cluster
whose density is 200 times higher than the critical density, is close to virial
radius and can be estimated by the formula $R_{200} =
\sqrt{3}\sigma/(10H(z))$\,Mpc. The mass within $R_{200}$ is equal to
$M_{200} = 3G^{-1}R_{200}\sigma_{200}^{2}$. We thus first estimate the average
radial velocity $cz$ of the cluster and its dispersion $\sigma$, and then
find the $R_{200}$ radius from the dispersion. We consider the
galaxies with velocities greater than $2.5\sigma$ to be background objects. We
use the iteration method to find all the cluster parameters within the
given radius.

We summed up the resulting masses of 10 clusters in UMa (with the exception of
A1291B and Anon2 clusters, which we discuss below) to infer a lower supercluster
mass estimate of  $2.25 \times 10^{15}$ $M_{\odot}$. The total mass of 11
clusters in the UMa neighborhood (without A1279) is equal to $1.70 \times 10^{15}$
$M_{\odot}$. Some
clusters of galaxies  (A1270, RXCJ1010, and RXJ1033) contain subclusters located
both within $R_{200}$ and near this radius, and therefore the inferred masses
are lower estimates. Table~\ref{data1:Kopylova_n} lists the derived
parameters with the errors corresponding to the errors of inferred $\sigma$. The
cluster centers  usually coincide with the positions of the brightest
galaxies of the corresponding clusters and are located close the centers
of X-ray radiation (if detected) with the exception for some clusters discussed
below. Table~\ref{data1:Kopylova_n} also lists the \mbox{0.1--2.4~keV} X-ray
luminosities adopted from the BAX database \cite{Sadat:Kopylova_n}. The number
of galaxies, $N_z$, is equal to the number of galaxies observed
within $R_{200}$ from the cluster center.
\begin{table*}[tbp]
\setcaptionmargin{0mm} \onelinecaptionstrue
\captionstyle{flushleft}
\caption{Dynamic properties of clusters}
\label{data1:Kopylova_n}
\medskip
\begin{tabular}{l|c|c|c|c|c|c|c} \hline
Cluster& $\alpha$~~(J2000)~~$\delta$& $cz_h$& $N_z$& $\sigma_c$& $R_{200}$&
$M_{200}$& $L_{0.1-2.4~keV}$\\
    &hh mm ss dd mm ss & km/s&  & km/s& Mpc& $10^{14}~M_{\odot}$&
$10^{44}$~erg/s\\
\hline
A1270    & 11 29 42.0+54 05 56 & 20688& 57& $553\pm73$& 1.32&
$2.82\pm1.06$&0.06:\\
A1291A   & 11 32 21.1+55 58 03 & 15394& 33& $391\pm68$& 0.94& $1.01\pm0.54$&0.22
\\
A1291B   & 11 32 02.4+56 04 12 & 17357& 37& $550\pm90$& 1.32& $2.80\pm1.39$& -
\\
A1318    & 11 36 03.5+55 04 31 & 16914& 40& $394\pm62$& 0.95&
$1.03\pm0.49$&0.04:\\
A1377    & 11 47 21.3+55 43 49 & 15531& 86& $632\pm68$& 1.53& $4.28\pm1.38$&0.28
\\
A1383    & 11 48 05.8+54 38 47 & 17862& 52& $464\pm64$& 1.12&
$1.69\pm0.74$&0.13:\\
A1436    & 12 00 08.8+56 10 52 & 19499& 89& $682\pm72$& 1.64&
$5.34\pm1.70$&0.52\\
Anon1    & 11 15 23.8+54 26 39 & 20951& 55& $608\pm82$& 1.46&
$3.78\pm1.53$&0.35\\
Anon2    & 11 19 46.0+54 28 02 & 21147& 14& $253\pm68$& 0.61& $0.27\pm0.22$& -\\
Anon3    & 11 29 32.3+55 25 20 & 20390& 23& $375\pm78$& 0.90& $0.88\pm0.56$& -\\
Anon4    & 11 39 08.5+55 39 52 & 18303& 25& $397\pm79$& 0.95& $1.05\pm0.64$& -\\
Sh166    & 12 03 11.9+54 50 50 & 15003& 24& $327\pm67$& 0.79& $0.59\pm0.38$& -\\
A1003    & 10 25 01.6+47 50 28 & 18882& 28&$562\pm106$&
1.35&$2.98\pm1.95$&0.10\\
A1169    & 11 07 49.3+43 55 00 & 17630& 69& $582\pm70$&
1.40&$3.32\pm1.20$&0.06:\\
A1279    & 11 31 39.3+67 14 30 & 16285&  6& $187\pm76$& 0.45&$0.11\pm0.14$& - \\
A1452    & 12 03 28.4+51 42 56 & 18542& 20&$514\pm115$& 1.23&$2.27\pm1.71$& -\\
A1461    & 12 04 24.7+42 33 43 & 16177& 13& $317\pm88$& 0.77&$0.54\pm0.46$& -\\
A1507    & 12 14 48.6+59 54 22 & 17978& 38& $432\pm70$&
1.04&$1.36\pm0.66$&0.07\\
A1534    & 12 24 42.8+61 28 15 & 20967& 18& $307\pm72$& 0.74&$0.49\pm0.35$& -\\
RXCJ1010 & 10 10 16.1+54 30 06 & 13736& 34& $418\pm72$&
1.01&$1.24\pm0.60$&0.02\\
RXJ1033  & 10 33 51.2+57 03 21 & 13671& 47& $413\pm60$&
1.00&$1.19\pm0.52$&0.01\\
RXCJ1053A& 10 54 11.2+54 50 18 & 21551& 49& $507\pm72$&
1.21&$2.18\pm0.94$&0.53\\
RXCJ1053B& 10 51 47.0+55 23 00 & 22113& 30& $420\pm77$& 1.00&$1.23\pm0.42$& -\\
RXCJ1122 & 11 22 45.8+67 09 55 & 16607& 12& $223\pm64$&
0.54&$0.19\pm0.16$&0.06\\
\hline
\end{tabular}
\end{table*}

\begin{figure*}[tbp]
\setcaptionmargin{0mm}
\onelinecaptionsfalse
\includegraphics[scale=0.53,angle=-90]{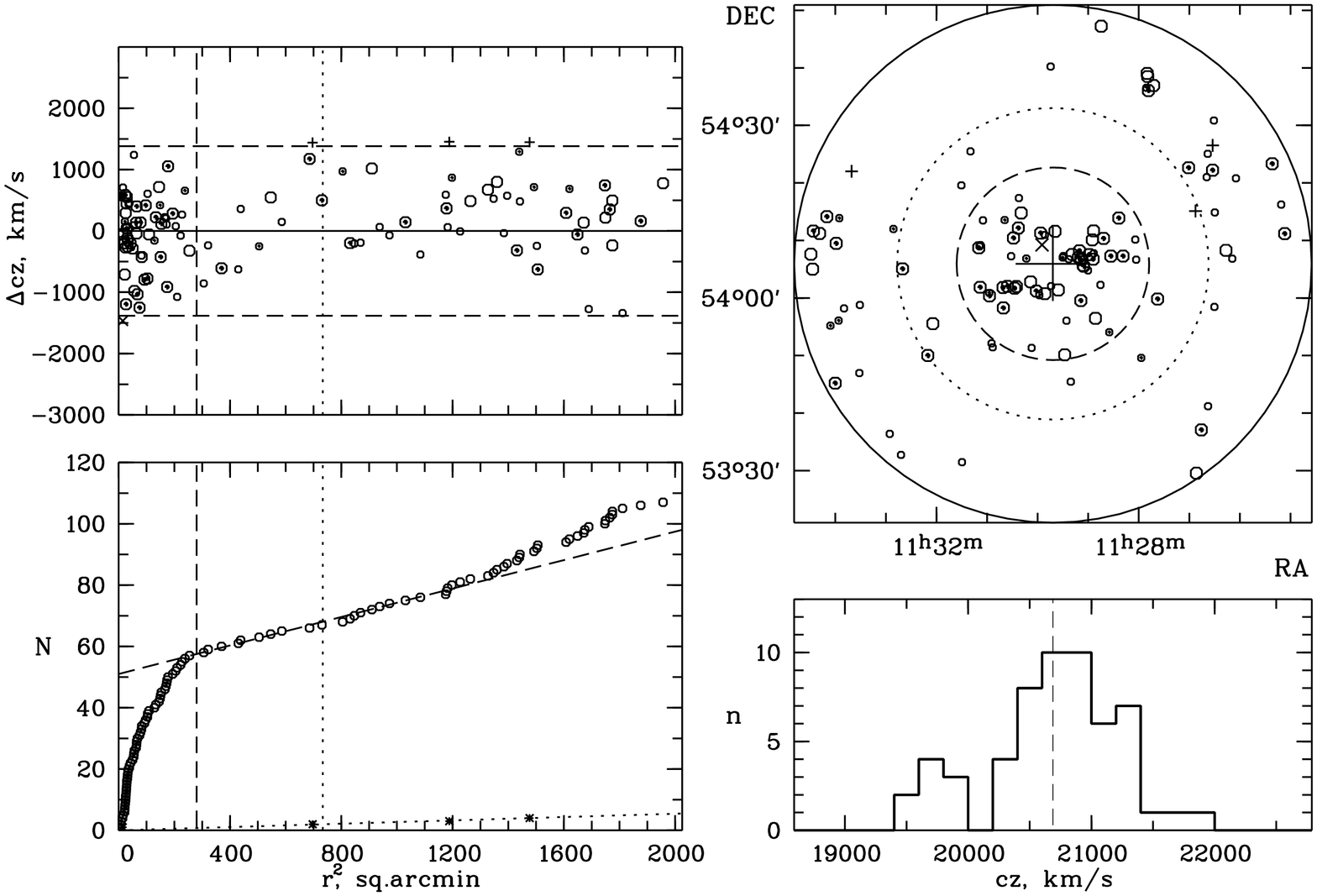}
\captionstyle{normal}
\caption{
Distribution of galaxies in A1270. The top left figure represents the deviation of
the radial velocities of galaxies from the mean radial velocity of the cluster,
averaged over the galaxies located within $R_{200}$ from the cluster center.
The horizontal dashed lines correspond to  $\pm2.5\sigma$ deviations; the
vertical dashed line indicates the $R_{200}$ radius, and the dotted line shows
the \mbox{Abell radius (2.14~Mpc).} The large circles indicate the galaxies
brighter than $M_K^*+1 = -23^m.29$; the circles with dots inside are the
early-type galaxies; plus signs and crosses are the background and foreground
galaxies, respectively. In the bottom left figure we present the integrated distribution
of the total number of galaxies as a function of the squared
clustercentric distance. The circles and asterisks correspond to the galaxies
shown by circles in the top left figure and to background galaxies,
respectively. The dashed and dotted lines indicate the domains of linear
increase of the number of cluster and background galaxies, respectively. In the top
right figure you can see the sky distribution (in equatorial coordinates) of the
galaxies represented in the top left figure (the same designations are used).
Centered circles denote the regions inside $R_{200}$ (the dashed curve) and
the Abell radius (the dotted curve). The domain under study is bounded by the
circle of radius 45$\arcmin$ (the solid line). The big cross indicates the
position of the center of the cluster. And finally in the bottom right figure we presented the
distribution of the radial velocities of the cluster galaxies located within
$R_{200}$. The vertical dashed line corresponds to the average radial
velocity of the cluster. The structure and designations are the same in
figs.~1--22.}
\label{clus1:Kopylova_n}
\end{figure*}

\begin{figure*}[tbp]
\setcaptionmargin{0mm}
\onelinecaptionsfalse
\includegraphics[scale=0.53,angle=-90]{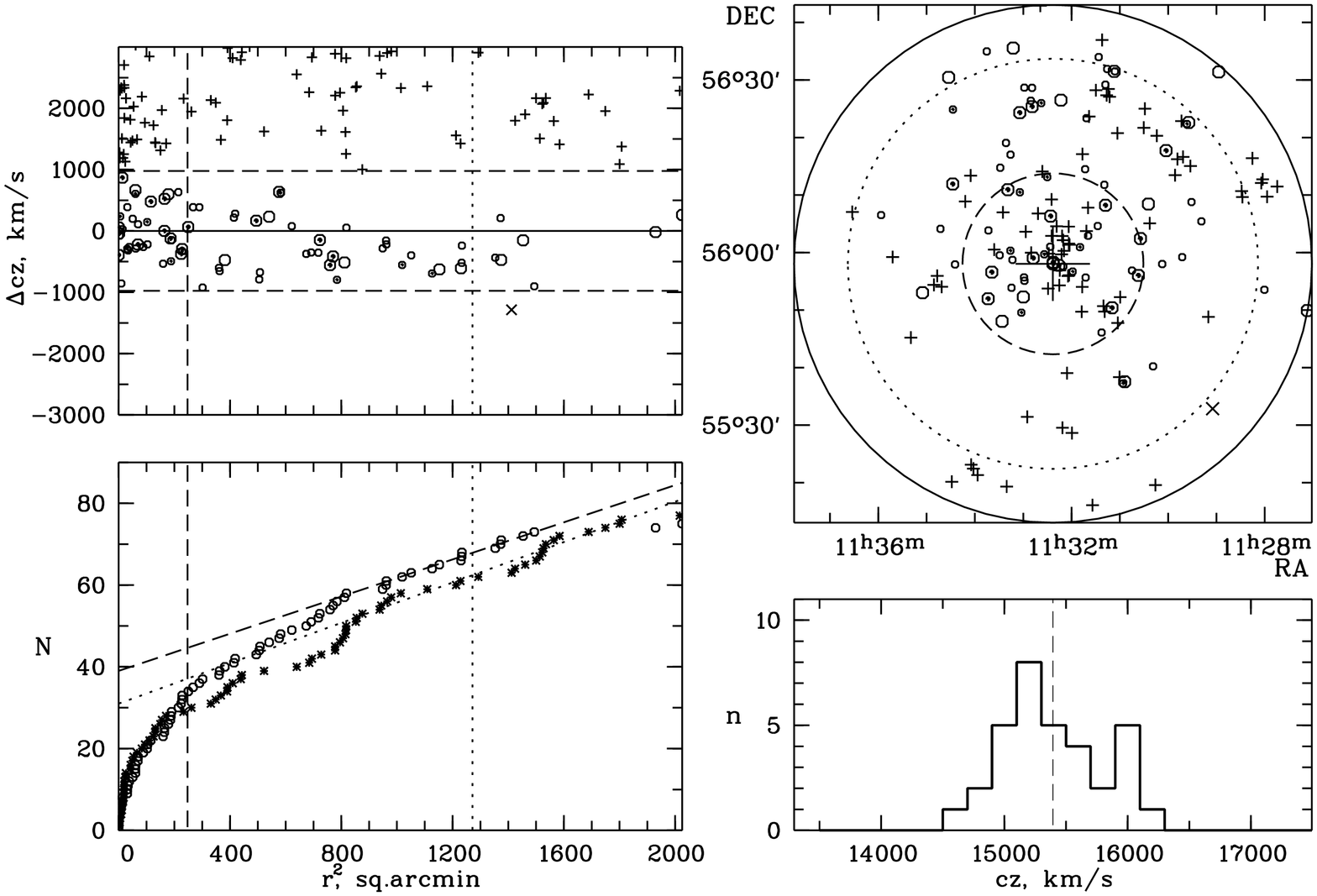}
\captionstyle{normal}
\caption{
Distribution of galaxies in A1291A.}
\label{clus2:Kopylova_n}
\includegraphics[scale=0.53,angle=-90]{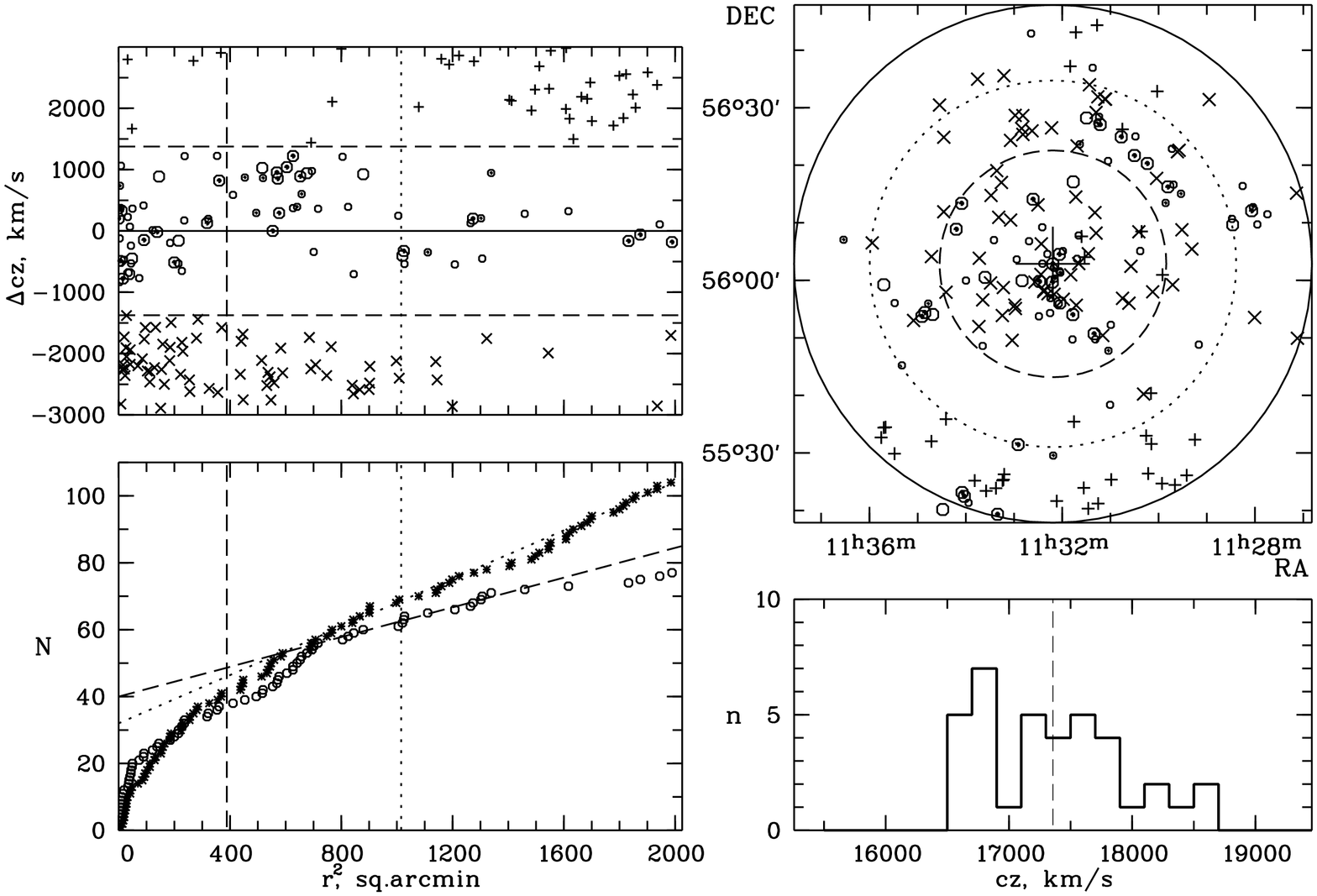}
\captionstyle{normal}
\caption{
Distribution of galaxies in A1291B.}
\label{clus3:Kopylova_n}
\end{figure*}

\begin{figure*}[tbp]
\setcaptionmargin{0mm}
\onelinecaptionsfalse
\includegraphics[scale=0.53,angle=-90]{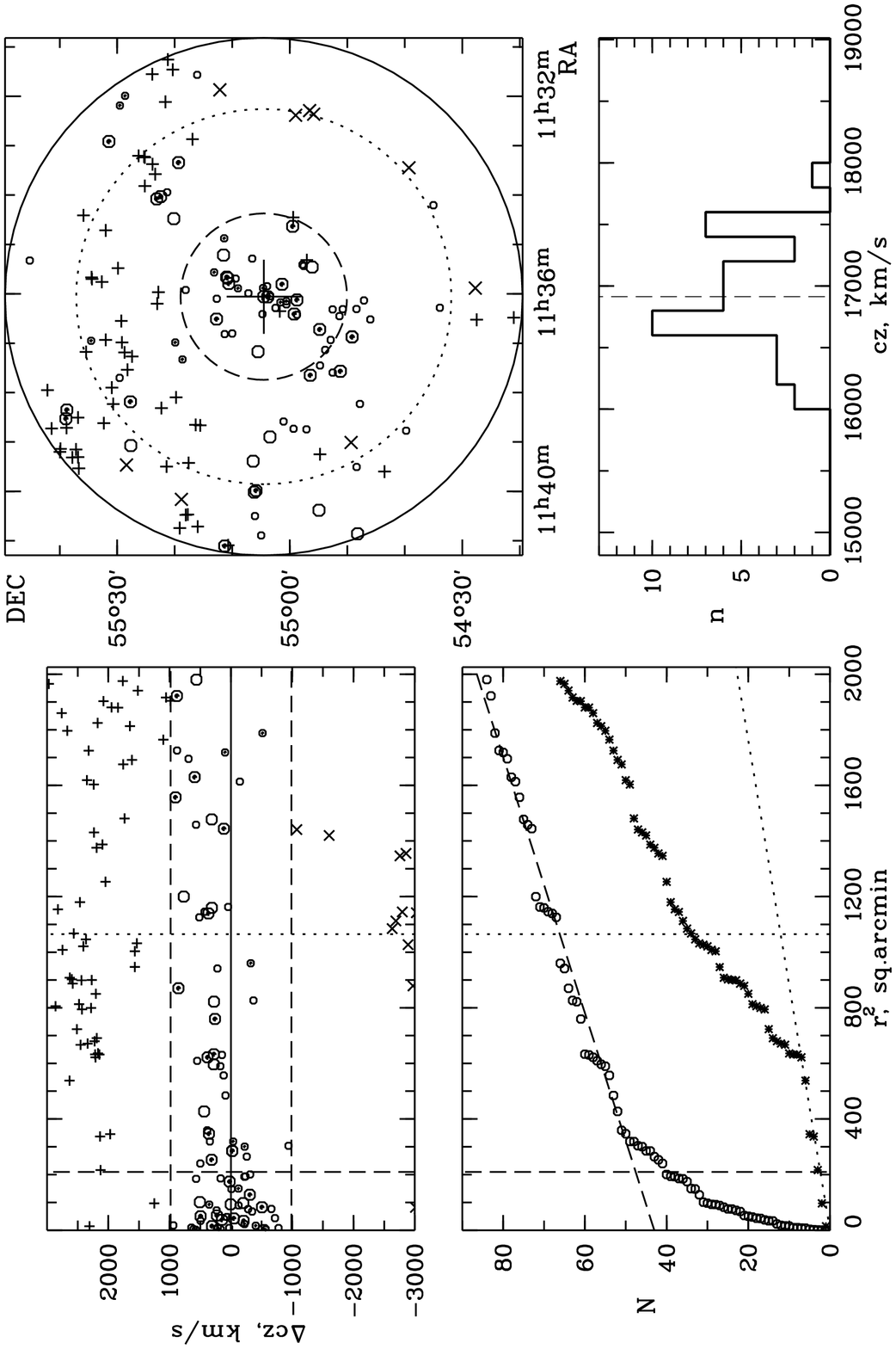}
\captionstyle{normal}
\caption{
Distribution of galaxies in A1318.}
\label{clus4:Kopylova_n}
\includegraphics[scale=0.53,angle=-90]{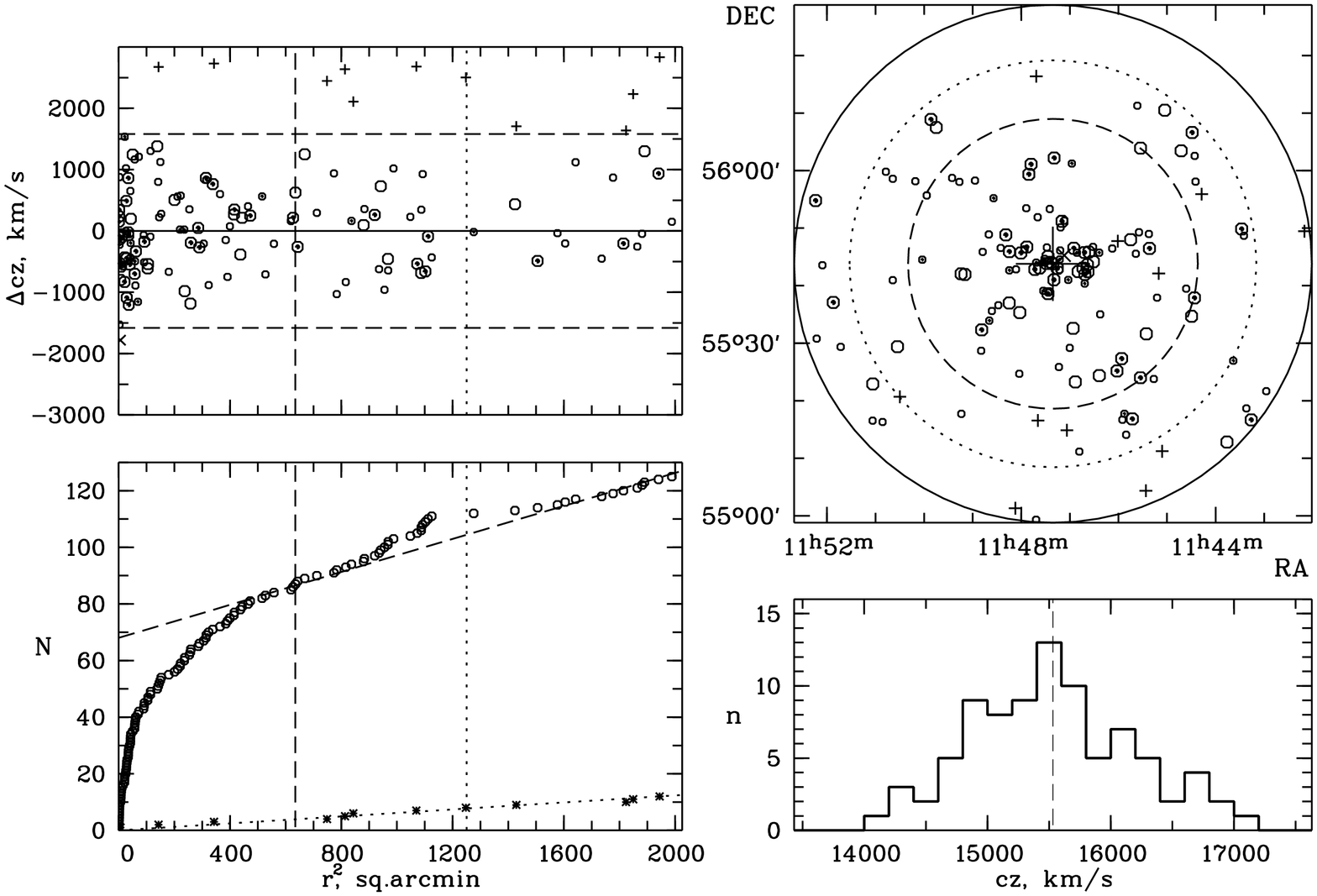}
\captionstyle{normal}
\caption{
Distribution of galaxies in A1377.}
\label{clus5:Kopylova_n}
\end{figure*}

\begin{figure*}[tbp]
\setcaptionmargin{0mm}
\onelinecaptionsfalse
\includegraphics[scale=0.53,angle=-90]{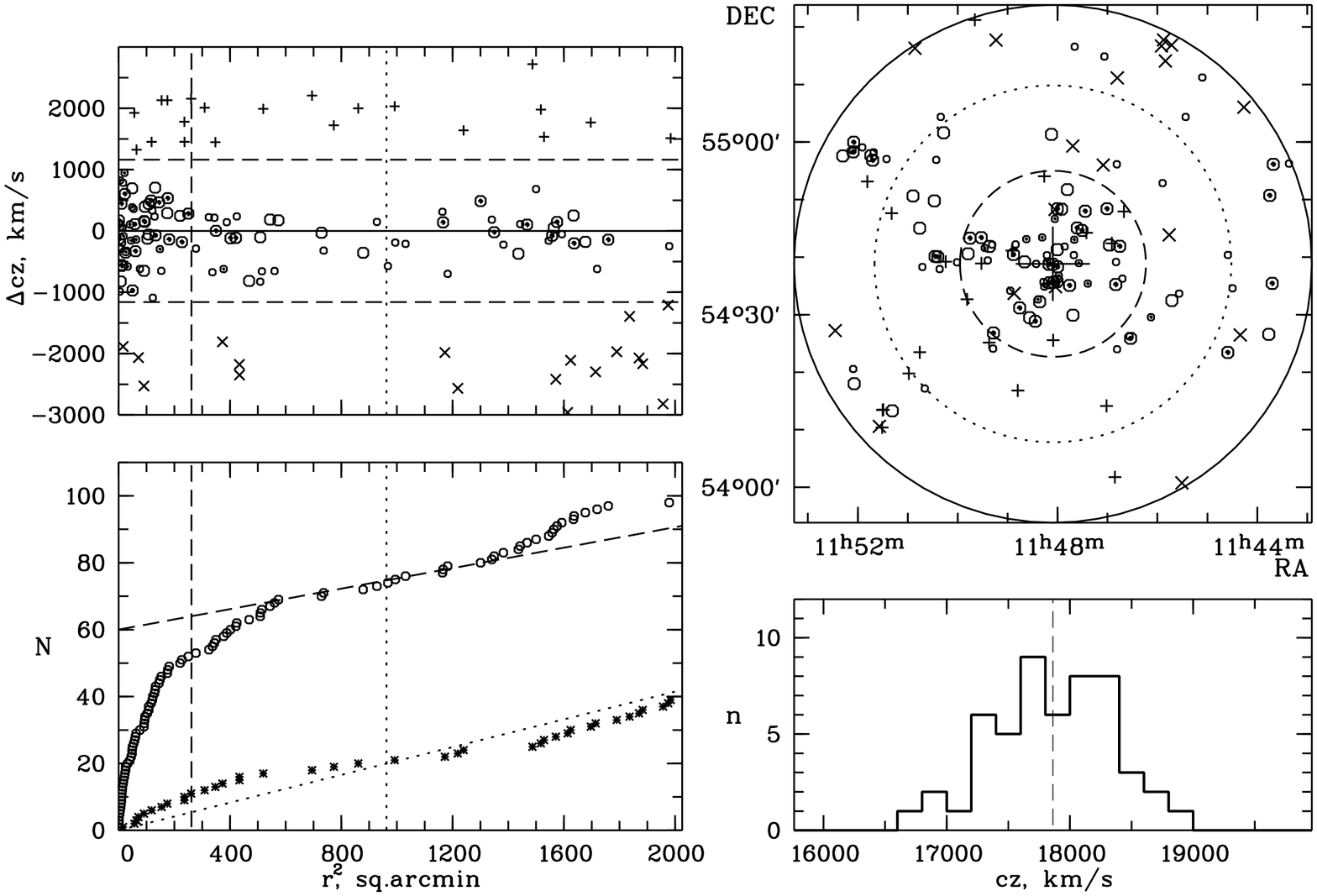}
\captionstyle{normal}
\caption{
Distribution of galaxies in A1383.}
\label{clus6:Kopylova_n}
\includegraphics[scale=0.53,angle=-90]{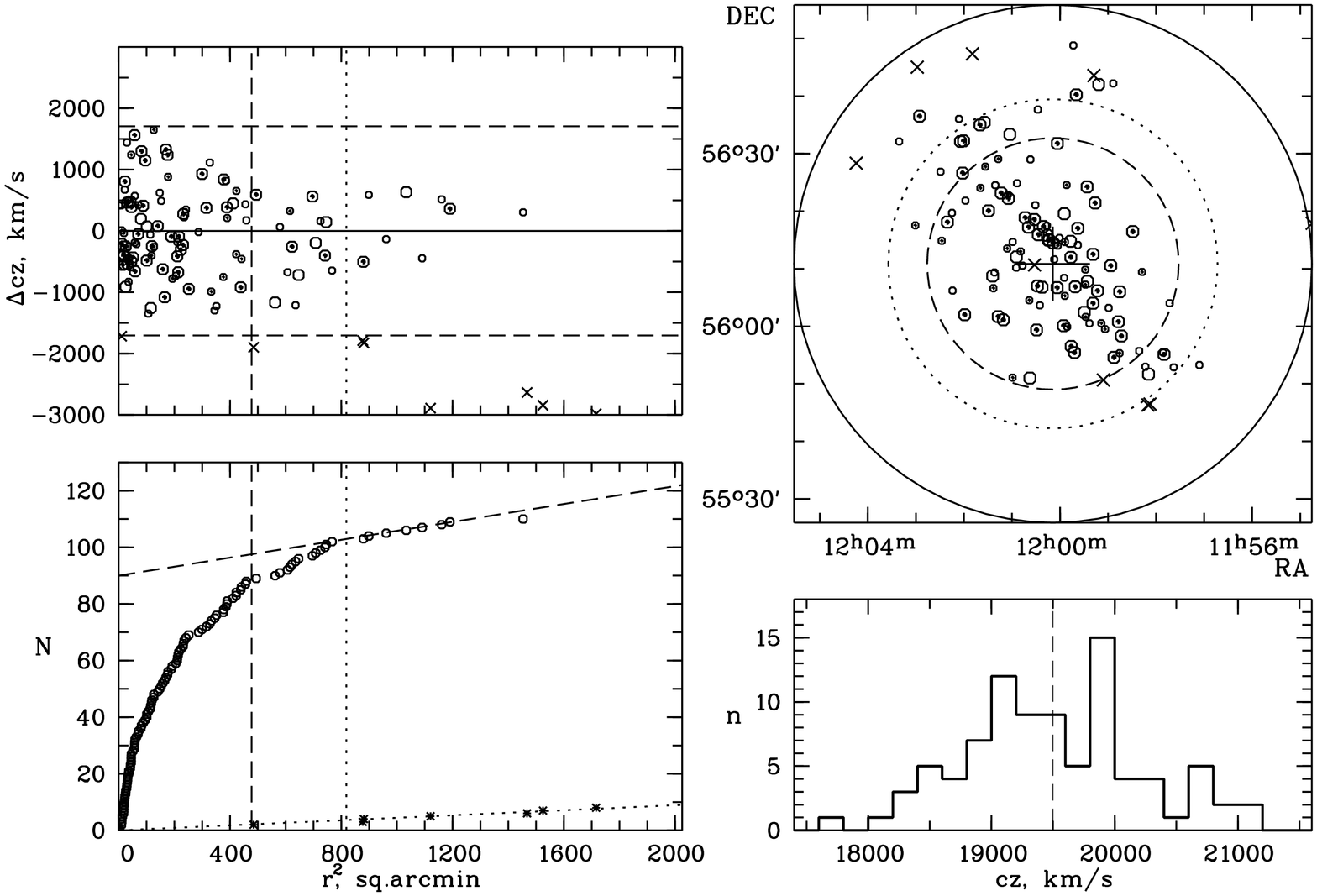}
\captionstyle{normal}
\caption{
Distribution of galaxies in A1436.}
\label{clus7:Kopylova_n}
\end{figure*}

In figs~\ref{clus1:Kopylova_n}--\ref{clus22:Kopylova_n} we present the following
properties of the clusters: the deviations of the radial velocities of galaxies
from the average radial velocity of the cluster within $\pm 3000$\,km/s; the
integrated distribution of the number of galaxies as a function of the squared
angular distance from the center of the cluster; the sky positions of galaxies
within  45$\arcmin$ from the center of the cluster, and the distribution of the
radial velocities of cluster member galaxies  located within $R_{200}$ from the
cluster center.

We compare our radial-velocity dispersions with those reported by Aguerri et
al.~\cite{Aguerri:Kopylova_n} for the nine clusters in common to find that the
differences are lying in the interval  ($-$53\,km/s)$<\Delta \sigma<$(+55~km/s). The
differences for the A1291B, RXCJ1053A, and A1169 clusters are much greater and amount to (+170\,km/s),
(+158\,km/s), and \mbox{($-$149\,km/s),} respectively; they
must be due to the presence of subclusters in these systems and to the
differences in the techniques used to find the position of the cluster
center and to identify cluster members.

\subsection{Comments on Some Clusters}

{\bf A1270}. This cluster is located in the farthermost filament of the UMa
system 
(see [5, Fig. 1c]).
It is evident from
Fig.~\ref{clus1:Kopylova_n}  that the cluster has emerged from the filament,
however, there are still subsystems within $R_{200}$. The
cluster does not have the brightest central galaxy and we adopt the midpoint
between the the two main subsystems as the position of the cluster center.

{\bf A1291A,B}. The A1291A cluster (Fig.~\ref{clus2:Kopylova_n}) has a
dominating galaxy. We adopted its position as the position of the cluster
center, although the center of X-ray radiation is slightly offset
with respect to it. The A1291B cluster deviates significantly from the relations
derived below (see Section~5), and at least two subsystems are apparent in its
radial-velocity distribution (Fig.~\ref{clus3:Kopylova_n})
(the bottom right figure). A plausible explanation would be to assume that the
A1291B cluster is actually a projection of several groups of galaxies oriented
along the line of sigh. In this case the dispersion of galaxy velocities and the
mass of the cluster should be highly overestimated.

{\bf A1436}. The cluster has an elongated shape and sharp edges. We may be
observing it during a period of violent dynamical relaxation after the merger of
two subclusters along the filament, when most of the galaxies have already
concentrated inside the virial radius, but the brightest galaxy had
not yet settled at the center of the cluster. As the center of the cluster we
adopted the position given by the mean coordinates of the two brightest
galaxies.

\begin{figure*}[tbp]
\setcaptionmargin{0mm}
\onelinecaptionsfalse
\includegraphics[scale=0.53,angle=-90]{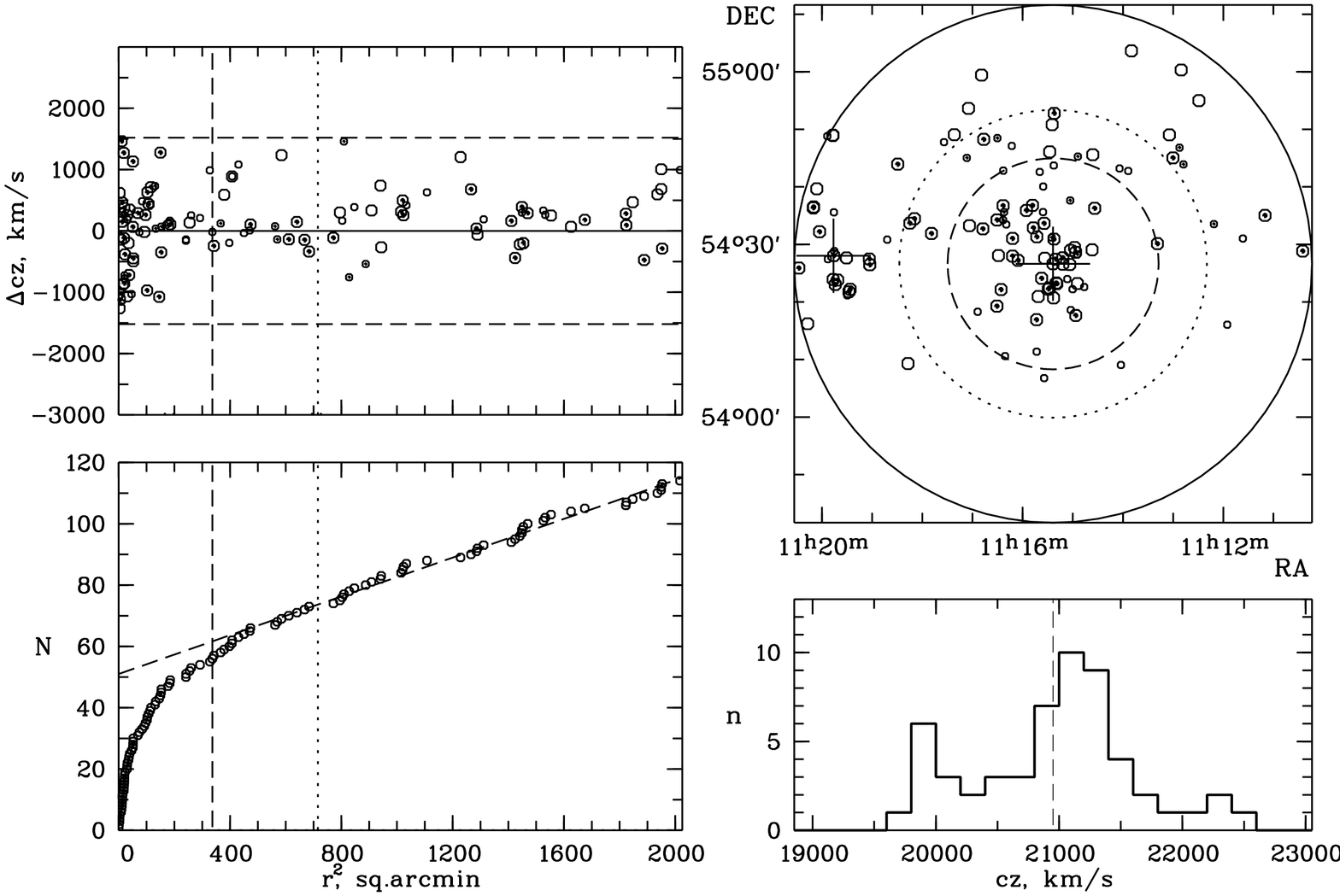}
\captionstyle{normal}
\caption{Distribution of galaxies in the  Anon1 cluster (the Anon2 cluster is
located left of Anon1 and we indicate the position of its center by a big
cross).}
\label{clus8:Kopylova_n}
\includegraphics[scale=0.53,angle=-90]{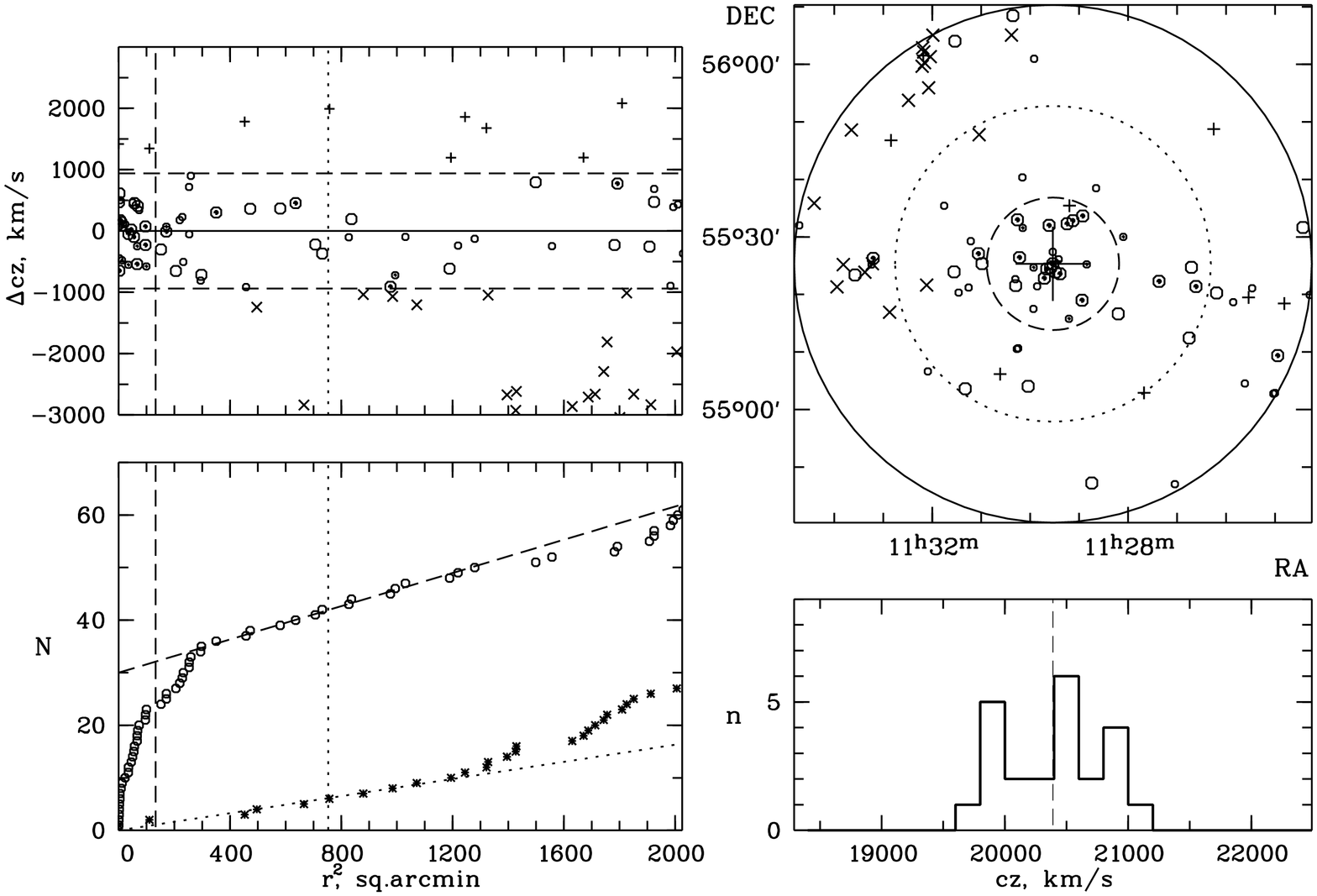}
\captionstyle{normal}
\caption{
Distribution of galaxies in Anon3.}
\label{clus9:Kopylova_n}
\end{figure*}

\begin{figure*}[tbp]
\setcaptionmargin{0mm}
\onelinecaptionsfalse
\includegraphics[scale=0.53,angle=-90]{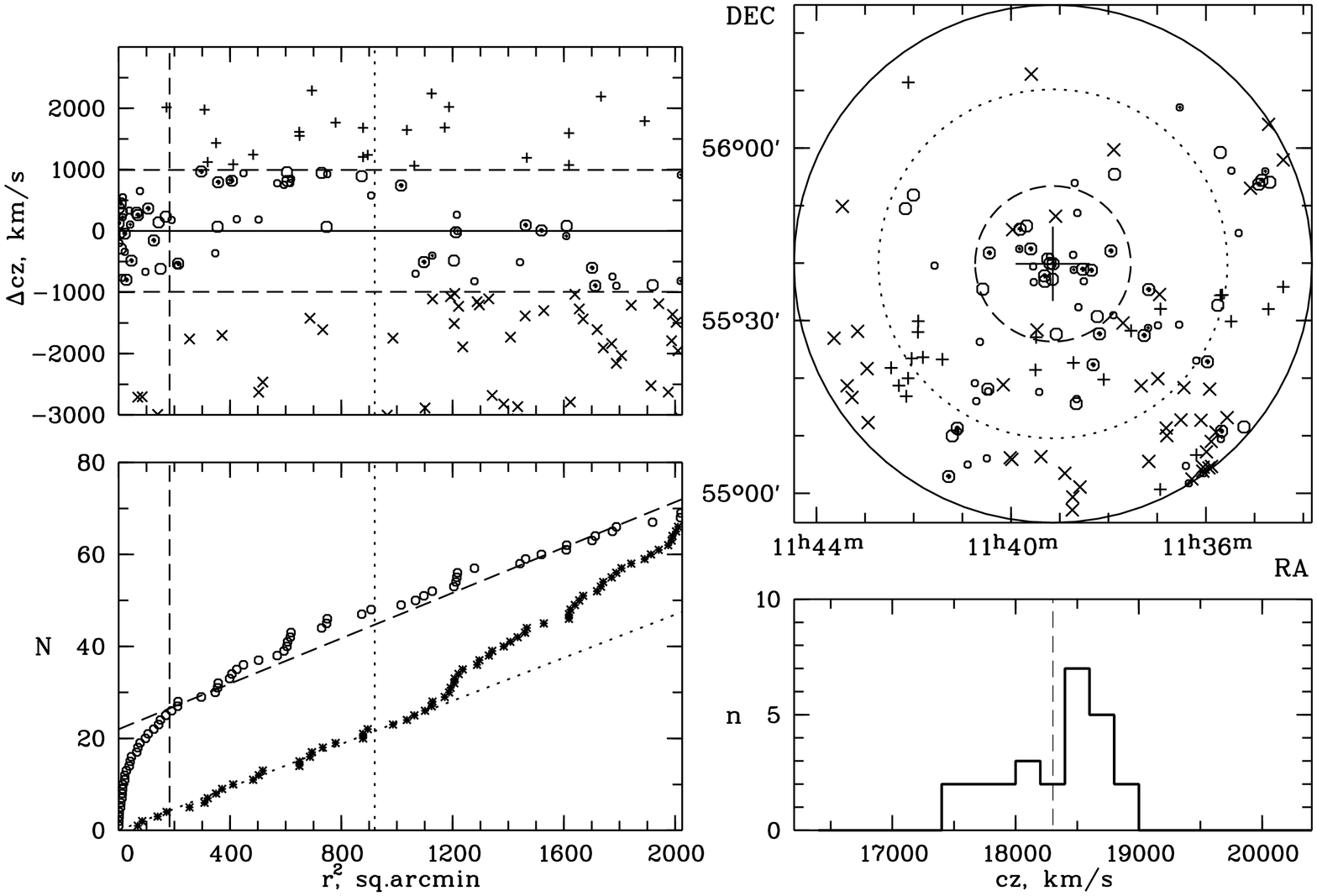}
\captionstyle{normal}
\caption{
Distribution of galaxies in Anon4.}
\label{clus10:Kopylova_n}
\includegraphics[scale=0.53,angle=-90]{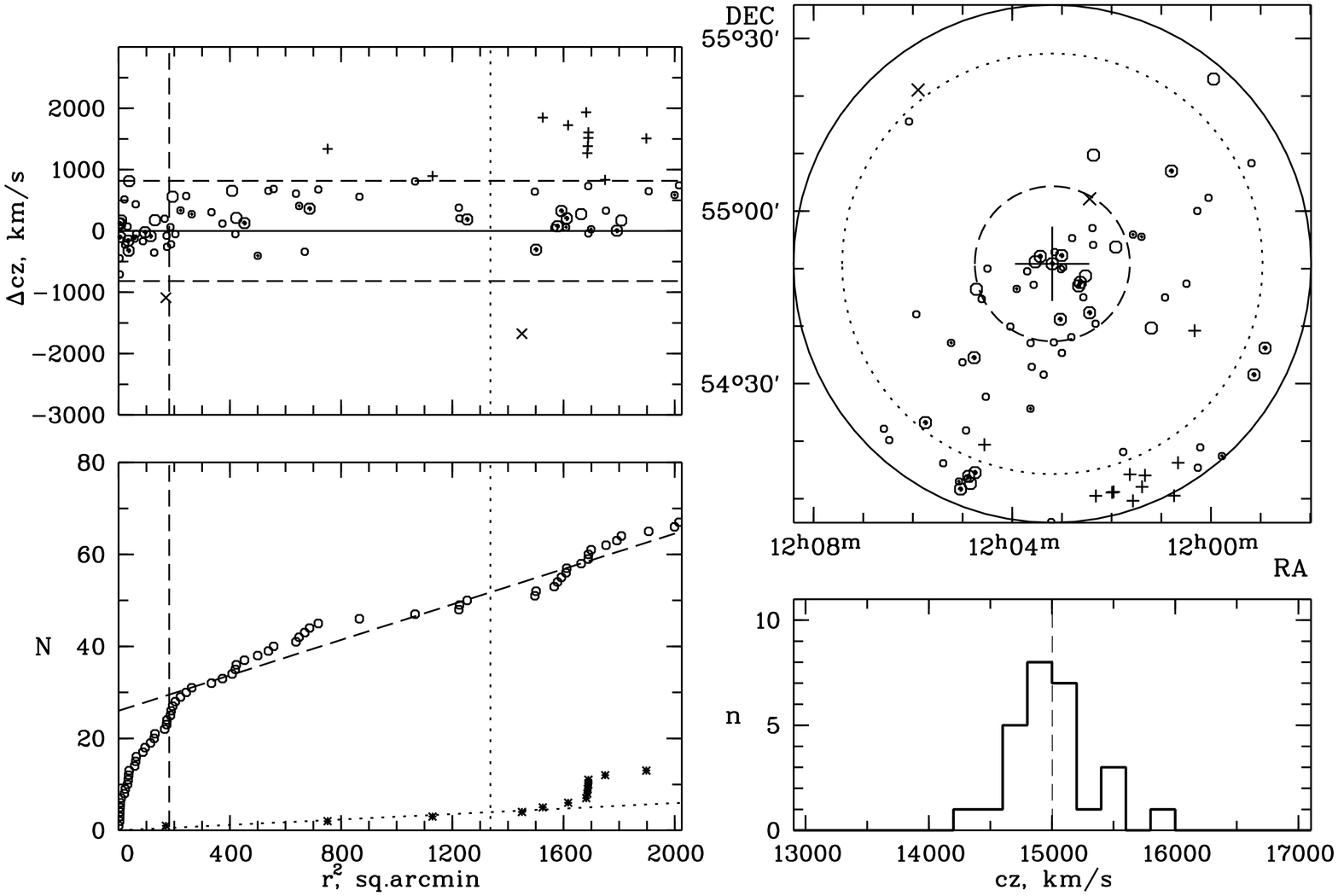}
\captionstyle{normal}
\caption{
Distribution of galaxies in Sh166.}
\label{clus11:Kopylova_n}
\end{figure*}

{\bf Anon1 (RXCJ1115.5+5426) and Anon2}.
These clusters are located in the farthermost filament of the UMa supercluster
(see [5, Fig. 1c]).
Anon1 (Fig.~\ref{clus8:Kopylova_n})
is a rather rich X-ray cluster with a cD galaxy at its center, which may be a
part of the filament adjacent to the Anon1 cluster rather than an isolated
virialized object. The Anon2 cluster, which is located $30\arcmin$ East of
Anon1, is a group consisting of bright early-type galaxies with a low dispersion
of radial velocities. That is why Anon2 does not obey the relations that we
derive for normal clusters (Section~5).

\begin{figure*}[tbp]
\setcaptionmargin{0mm}
\onelinecaptionsfalse
\includegraphics[scale=0.53,angle=-90]{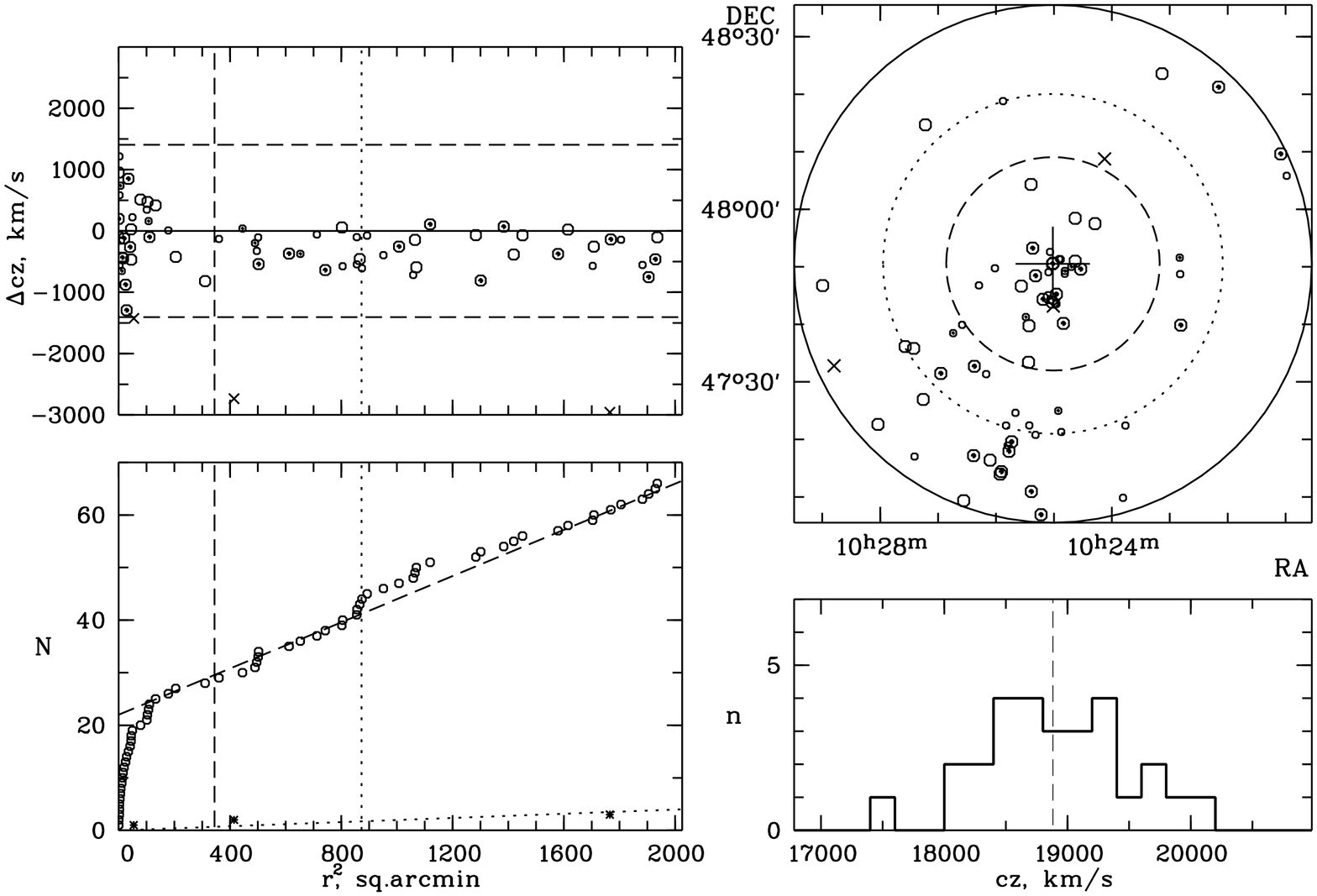}
\captionstyle{normal}
\caption{
Distribution of galaxies in  A1003.}
\label{clus12:Kopylova_n}
\includegraphics[scale=0.53,angle=-90]{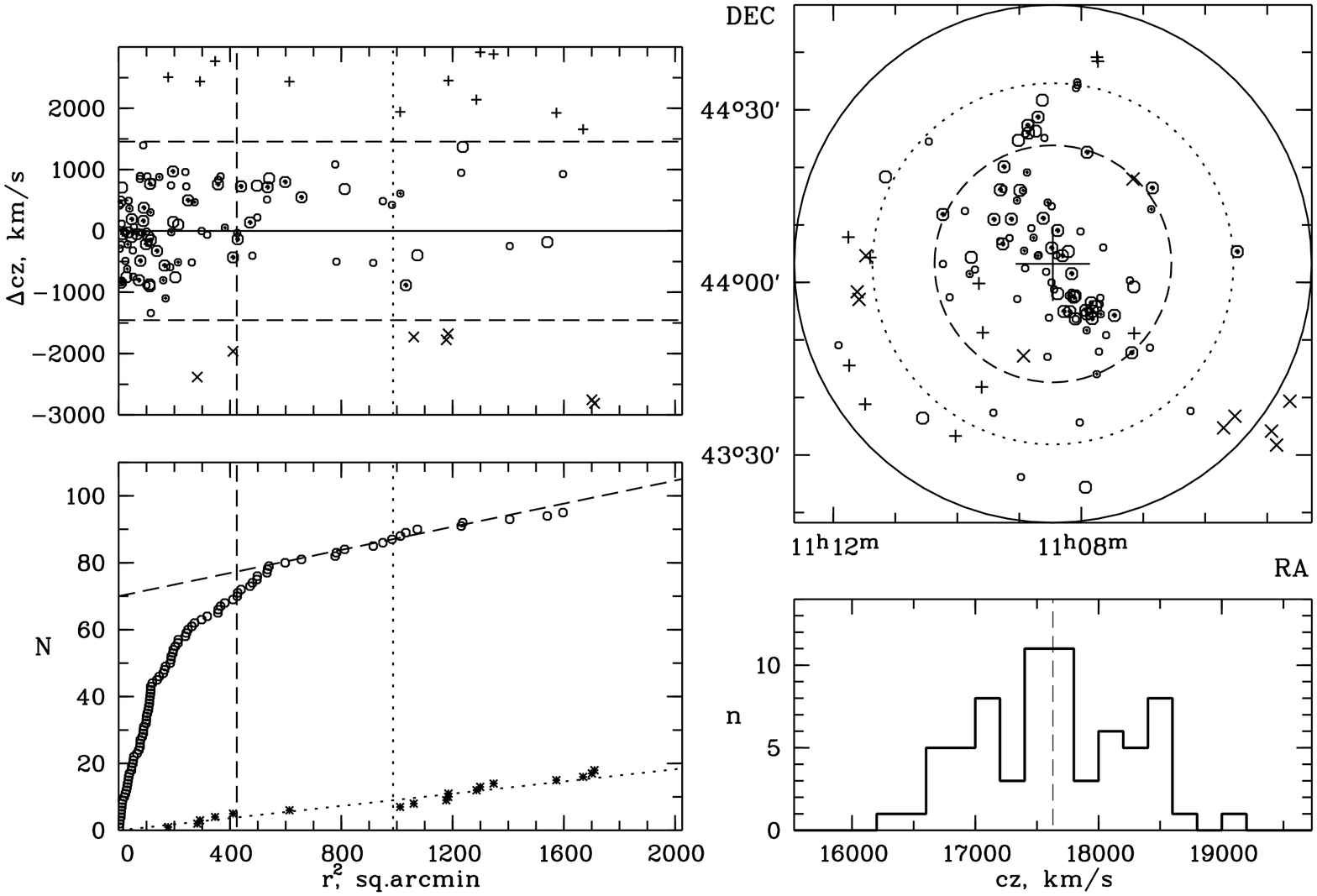}
\captionstyle{normal}
\caption{
Distribution of galaxies in A1169.}
\label{clus13:Kopylova_n}
\end{figure*}

{\bf A1169}. The cluster (Fig.~\ref{clus13:Kopylova_n}) is highly elongated in
the sky plane from Northeast to Southwest. As its center, we adopted the
centroid computed over all galaxies, although the brightest galaxy is located in
the Southwestern compact subgroup of galaxies.

\begin{figure*}[tbp]
\setcaptionmargin{0mm}
\onelinecaptionsfalse
\includegraphics[scale=0.53,angle=-90]{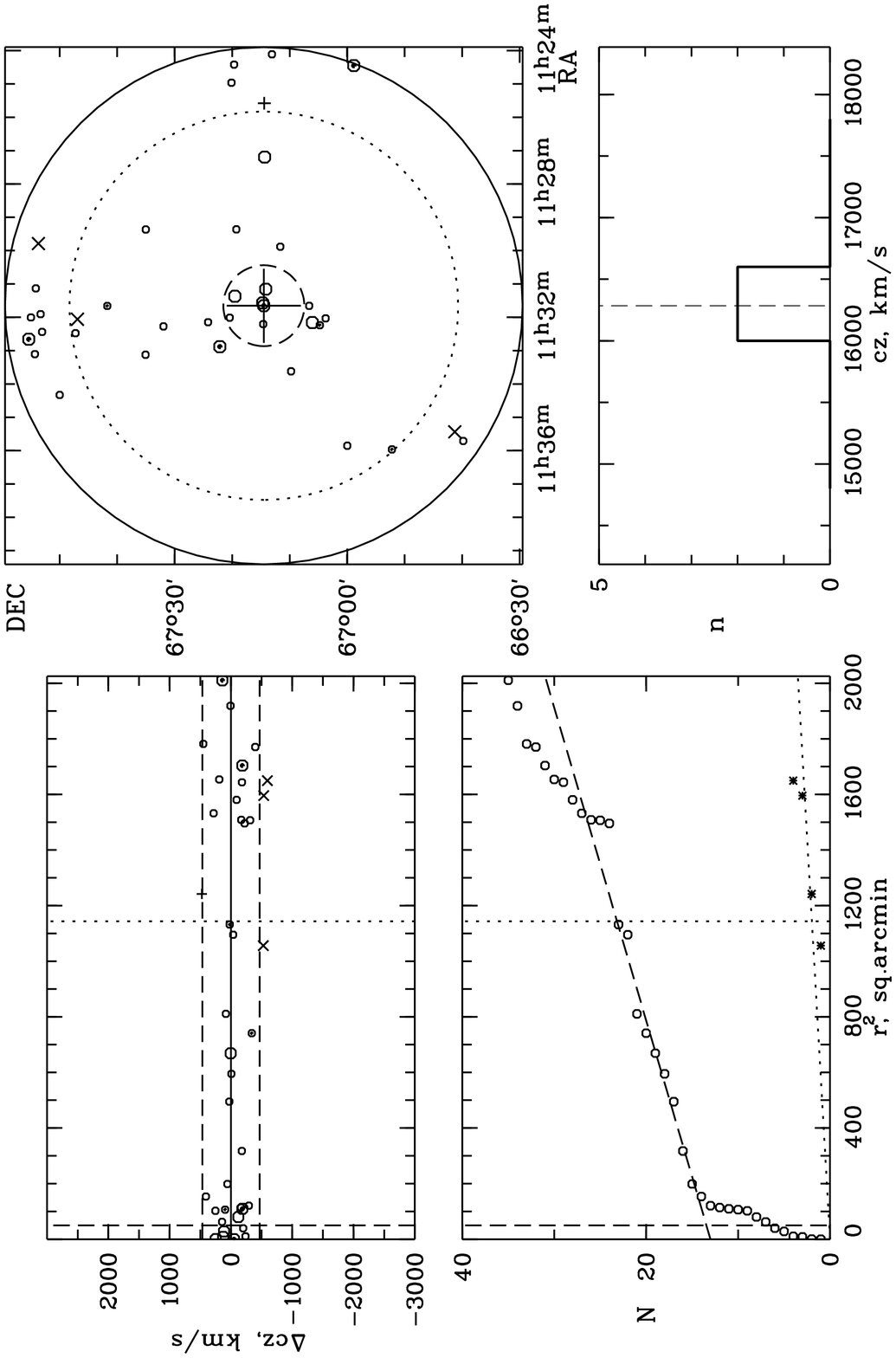}
\captionstyle{normal}
\caption{
Distribution of galaxies in A1279.}
\label{clus14:Kopylova_n}
\includegraphics[scale=0.53,angle=-90]{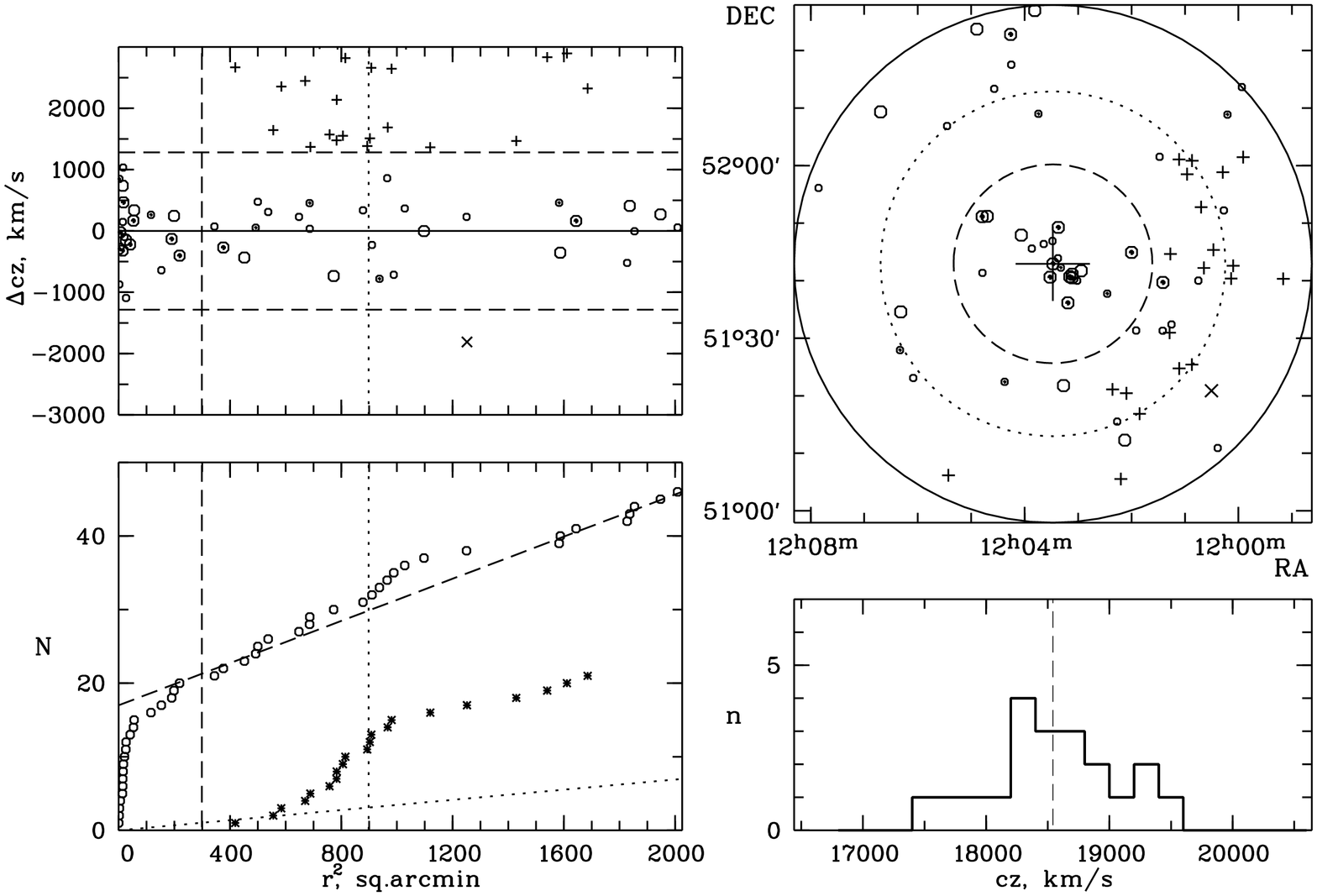}
\captionstyle{normal}
\caption{
Distribution of galaxies in A1452.}
\label{clus15:Kopylova_n}
\end{figure*}

\begin{figure*}[tbp]
\setcaptionmargin{0mm}
\onelinecaptionsfalse
\includegraphics[scale=0.53,angle=-90]{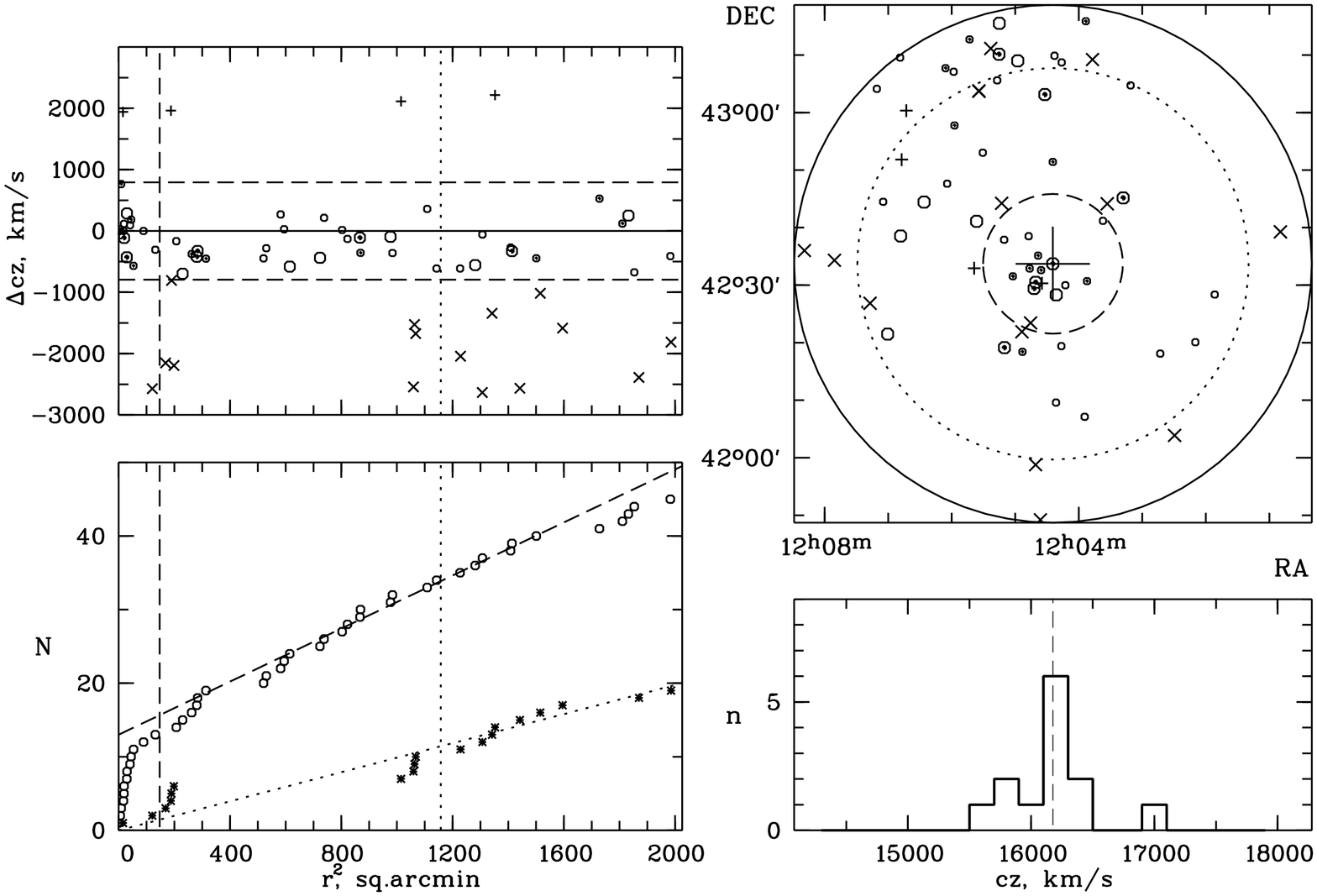}
\captionstyle{normal}
\caption{
Distribution of galaxies in  A1461.}
\label{clus16:Kopylova_n}
\includegraphics[scale=0.53,angle=-90]{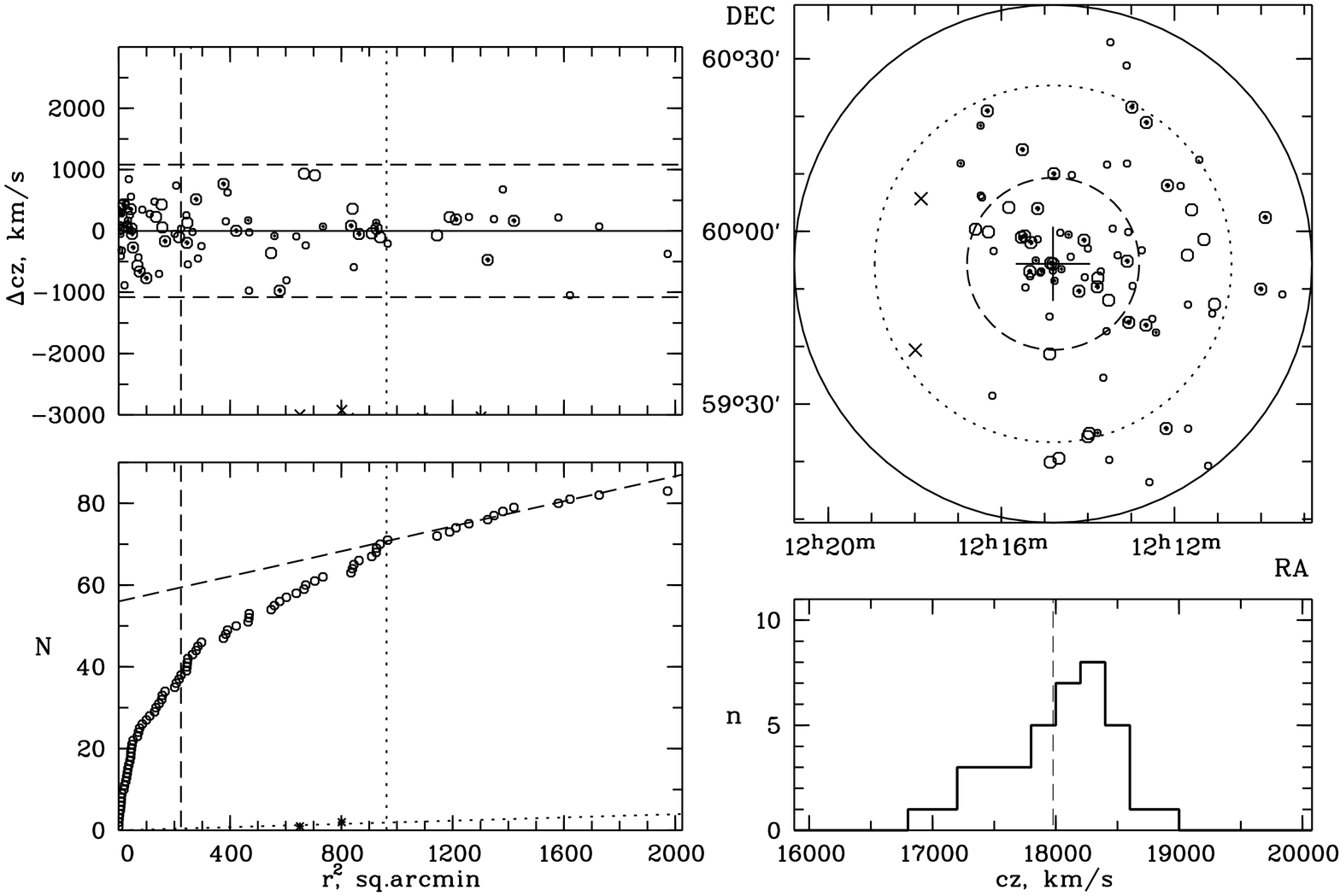}
\captionstyle{normal}
\caption{
Distribution of galaxies in A1507.}
\label{clus17:Kopylova_n}
\end{figure*}

{\bf A1279}. A very poor cluster (Fig.~\ref{clus14:Kopylova_n}) with one
elliptical galaxy and several late-type galaxies located within $R_{200}$. This
cluster has the lowest fraction (0.25) of early-type galaxies down to a limiting
magnitude of $M_K^*+1$. By its parameters, A1279 should be classified as a
group of galaxies.

\begin{figure*}[tbp]
\setcaptionmargin{0mm}
\onelinecaptionsfalse
\includegraphics[scale=0.53,angle=-90]{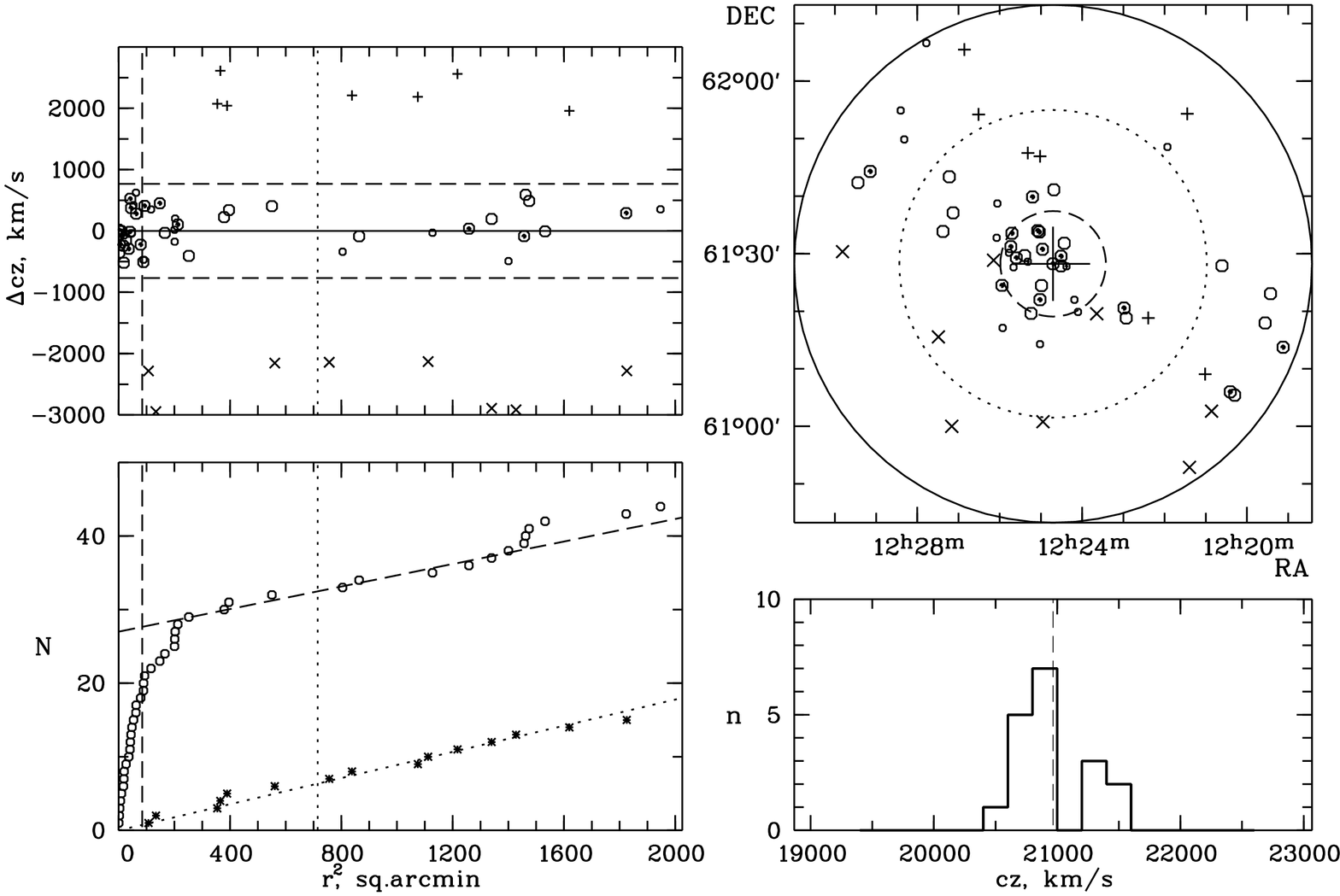}
\captionstyle{normal}
\caption{
Distribution of galaxies in A1534.}
\label{clus18:Kopylova_n}
\includegraphics[scale=0.53,angle=-90]{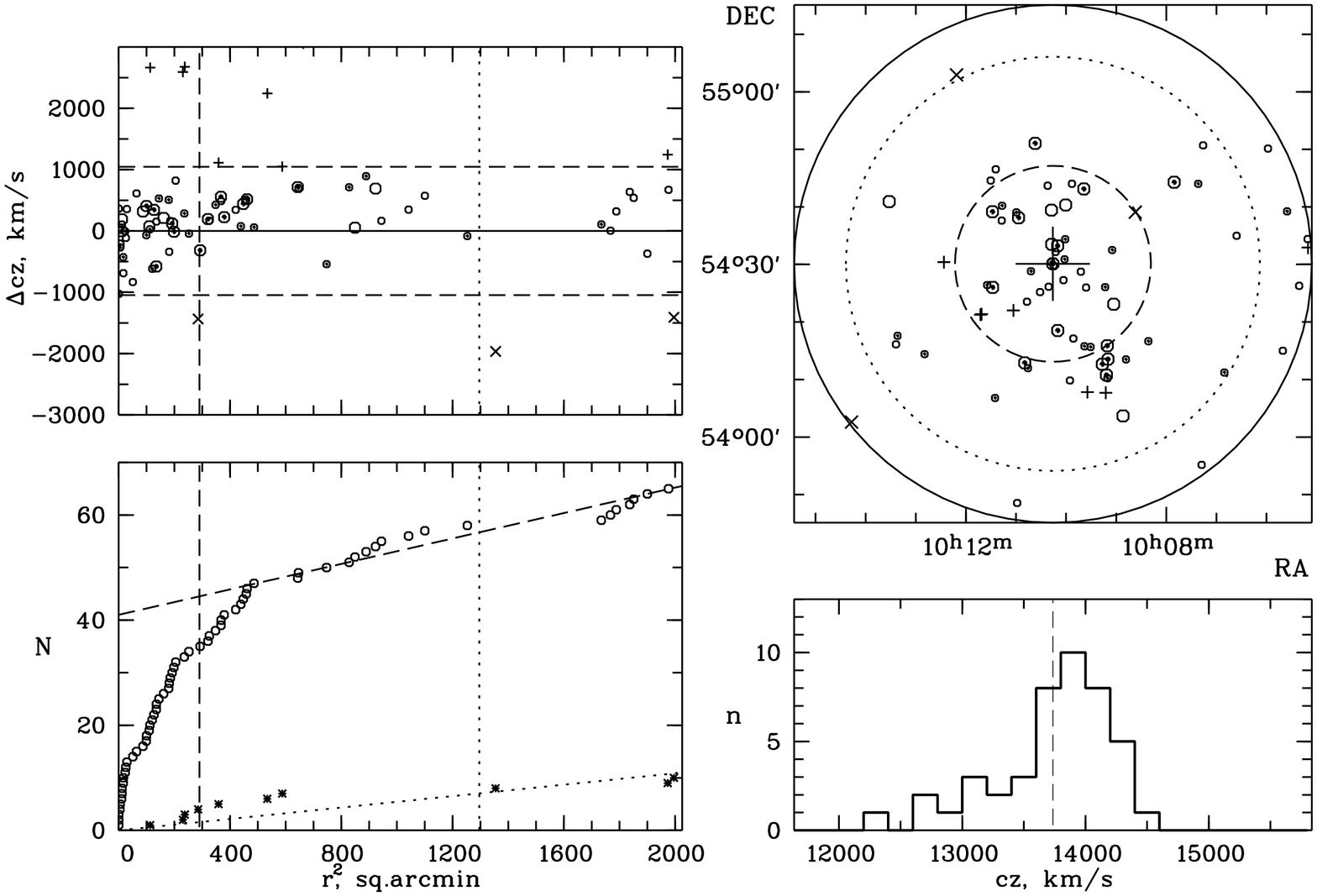}
\captionstyle{normal}
\caption{
Distribution of galaxies in RXCJ1010.}
\label{clus19:Kopylova_n}
\end{figure*}

{\bf RXCJ1010}. This cluster has not yet entirely formed
(Fig.~\ref{clus19:Kopylova_n}).
The center of the cluster is located at the center of X-ray flux distribution,
which coincides with the position of the brightest galaxy. It is evident from
Fig.~\ref{clus19:Kopylova_n} (the top right figure)
that the cluster contains small subclusters located near the center  (about
$9\arcmin$ and $18\arcmin$ from it), which consist of early-type galaxies.

\begin{figure*}[tbp]
\setcaptionmargin{0mm}
\onelinecaptionsfalse
\includegraphics[scale=0.53,angle=-90]{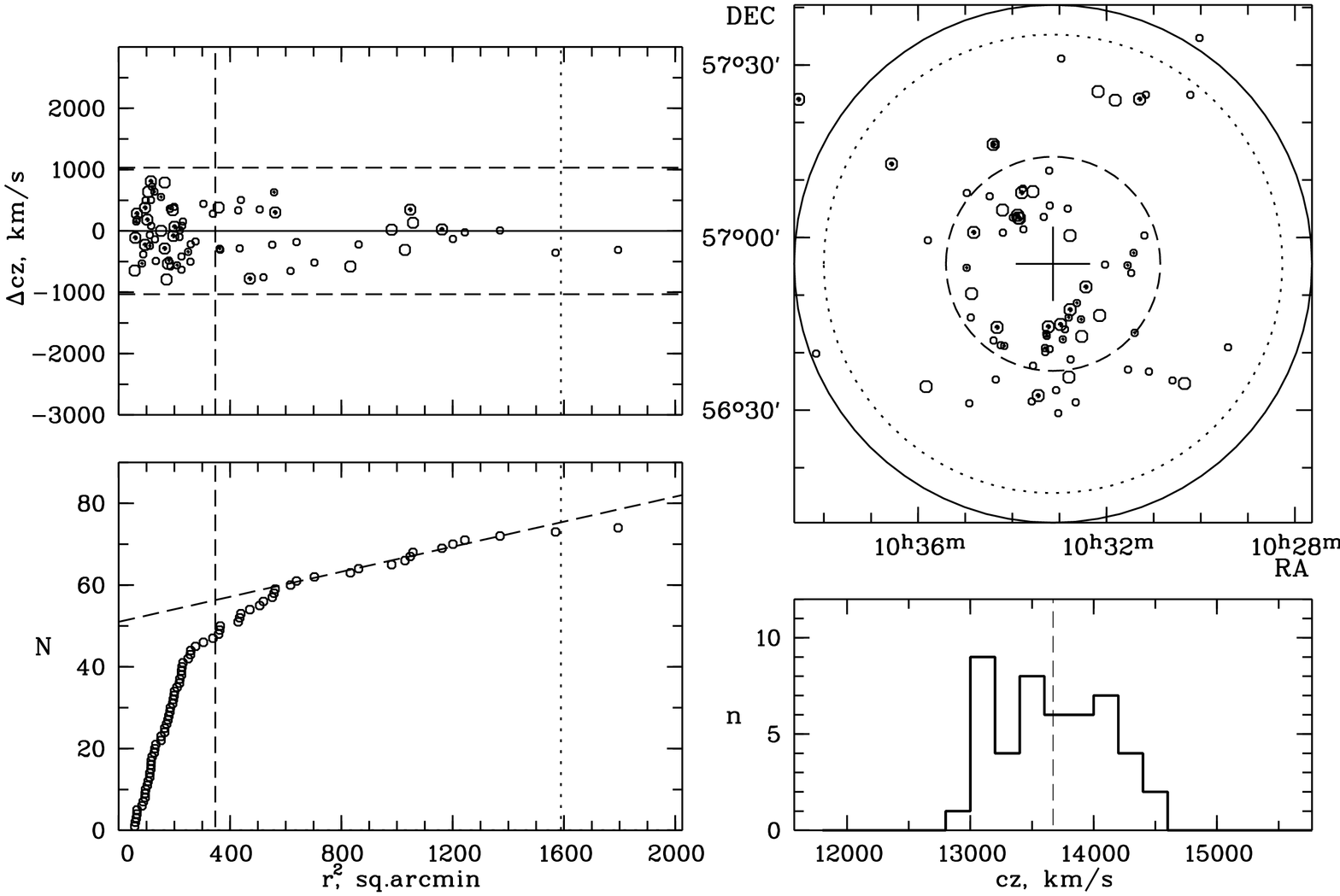}
\captionstyle{normal}
\caption{
Distribution of galaxies in RXJ1033.}
\label{clus20:Kopylova_n}
\includegraphics[scale=0.53,angle=-90]{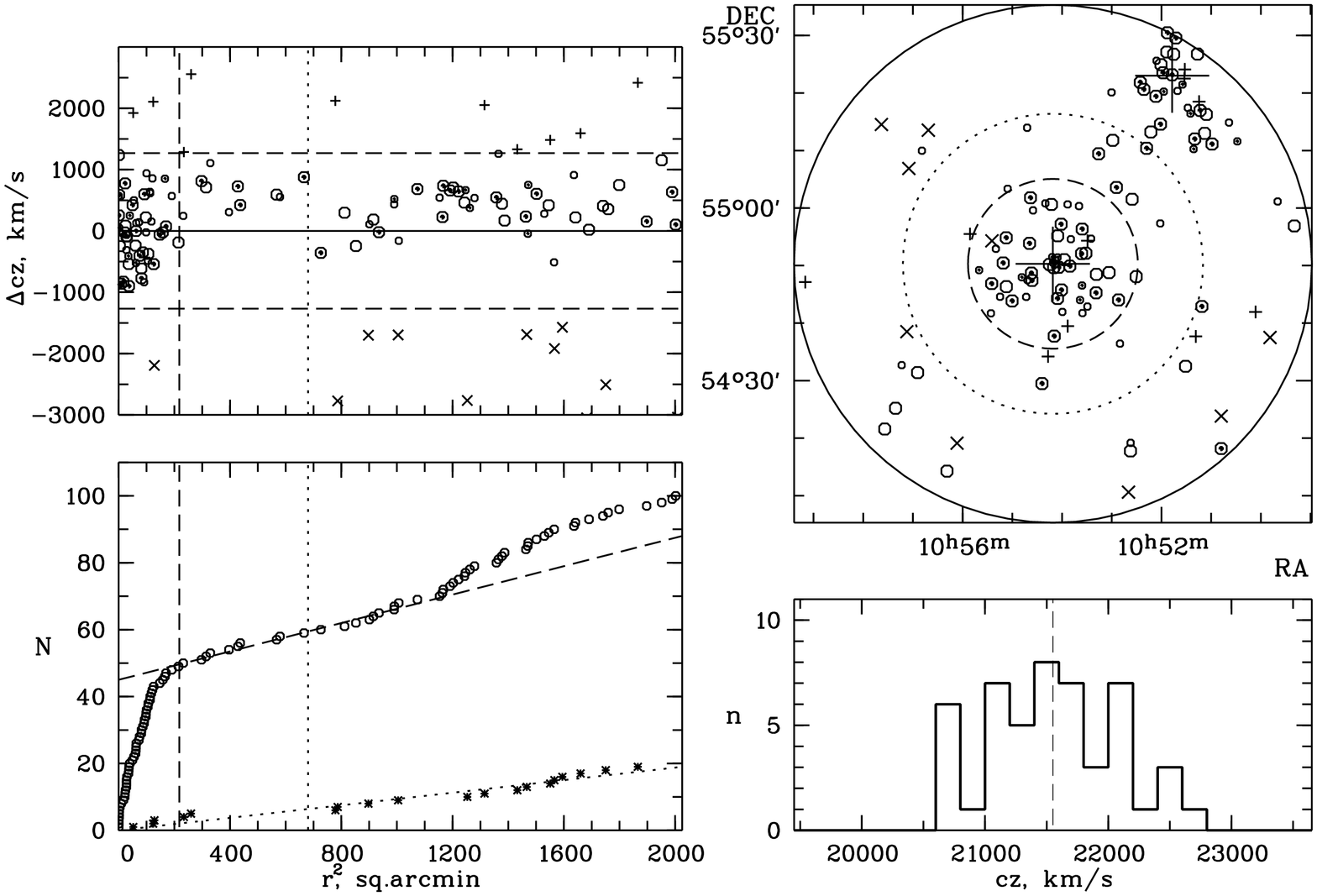}
\captionstyle{normal}
\caption{
Distribution of galaxies in RXCJ1053A (RXCJ1053B is located at the top right
corner and is indicated
by a big cross).}
\label{clus21:Kopylova_n}
\end{figure*}

{\bf RXJ1033}. The coordinates of the center of the X-ray flux distribution
approximately coincide with those of the brightest galaxy. However, given the
presence of another subcluster located  $15'$ from this center
(Fig.~\ref{clus20:Kopylova_n}), which consists mostly of late-type galaxies and
has a radial velocity that differs by  300\,km/s from that of the main cluster,
we assume that the center of the cluster should coincide with the position of
the centroid computed over all galaxies of the cluster.

{\bf RXCJ1053A,B}. There is another cluster (which we designate as RXCJ1053B)
$33\arcmin$ Northwest of the main cluster discovered via X-ray observations
(Fig.~\ref{clus21:Kopylova_n}). The average radial velocity of RXCJ1053B differs
by about  600\,km/s from that of the main cluster.

\begin{figure*}[tbp]
\setcaptionmargin{0mm}
\onelinecaptionsfalse
\includegraphics[scale=0.53,angle=-90]{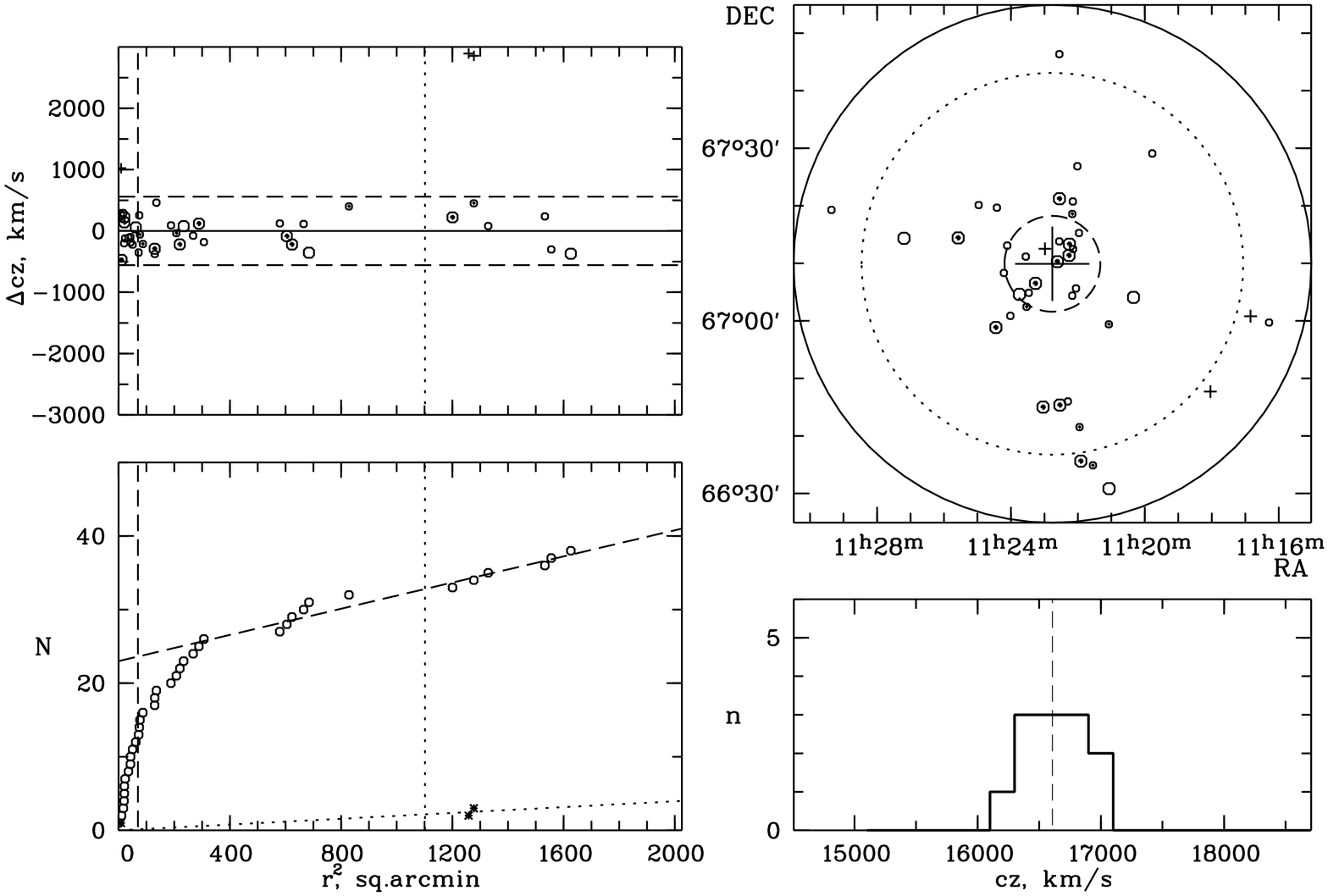}
\captionstyle{normal}
\caption{
Distribution of galaxies in RXCJ1122.}
\label{clus22:Kopylova_n}
\end{figure*}

{\bf RXCJ1122}. We set the center of the cluster (Fig.~\ref{clus22:Kopylova_n})
at the midpoint between the two brightest galaxies.

\section{TOTAL $K$-BAND LUMINOSITY OF CLUSTERS OF GALAXIES}
\begin{figure}[tbp]
\setcaptionmargin{0mm}
\onelinecaptionsfalse
\includegraphics[scale=0.4,angle=-90]{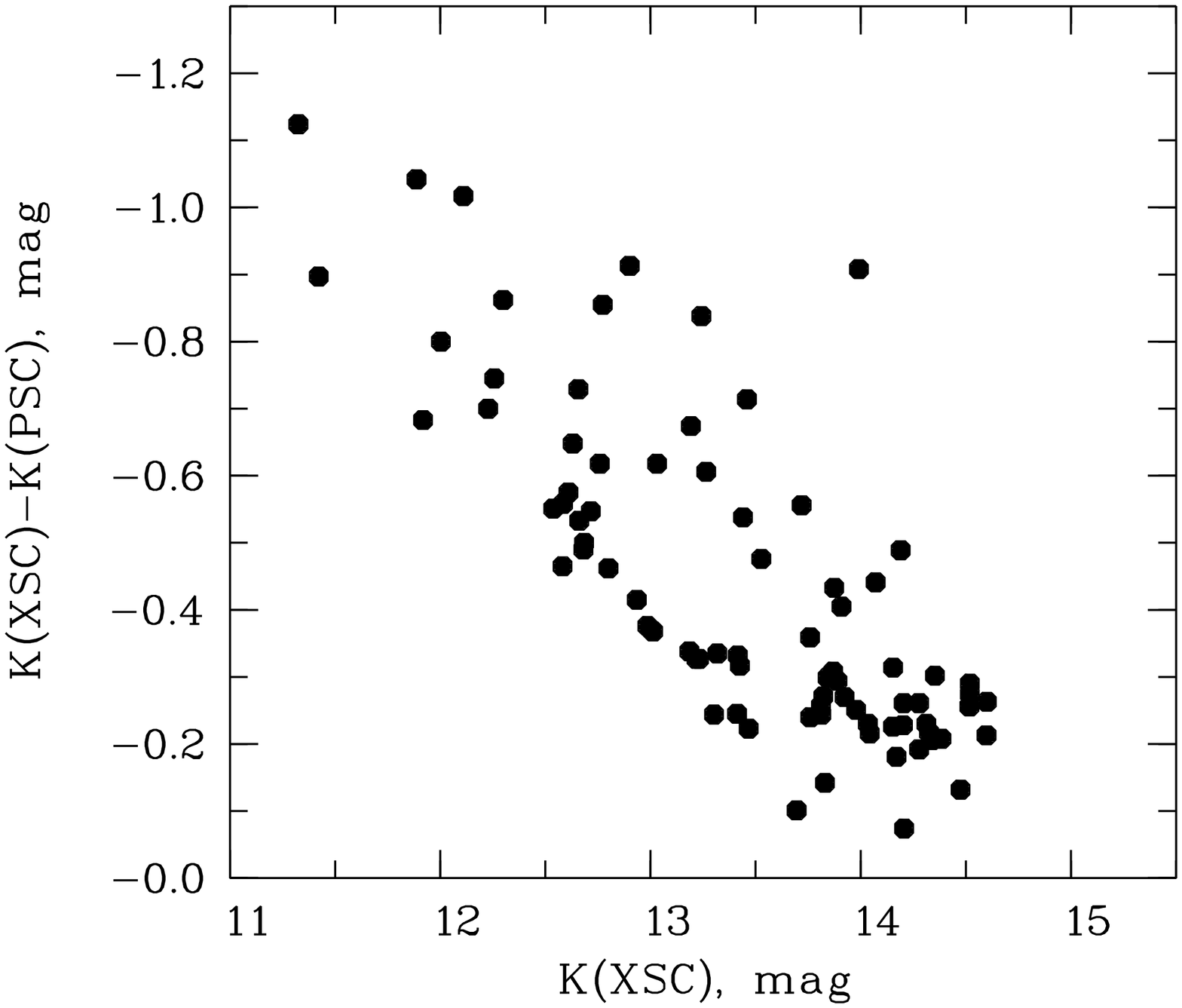}
\captionstyle{normal}
\caption{Comparison of the $K$-band magnitudes for the galaxies from the  2MASS
extended-source catalog XSC (magnitudes within the $\mu_K$~= 20$^m /\square''$
surface-brightness isophote) and the  2MASS point-source catalog PSC (magnitude
in the standard aperture) for the A1377 cluster.}
\label{point:Kopylova_n}
\end{figure}

The main baryon components of clusters of galaxies are: stars in galaxies and in
the intergalactic space and the hot gas that fills the intergalactic and
intercluster space. The parameters of these baryonic components can be measured
in the process of observations. Infrared (IR) radiation of stars is not affected
significantly by either starbursts in a galaxy or dust, because the central
regions of clusters of galaxies are mostly occupied  by early-type galaxies with
old stellar population. IR radiation is therefore a more accurate tracer of
mass of the stellar population in clusters of galaxies and it is often used for this
\mbox{(see, e.g., \cite{Lin:Kopylova_n,Ramella:Kopylova_n,Rines:Kopylova_n}).}
To find the total IR luminosities of clusters of galaxies, in  our earlier
paper \cite{Kop2:Kopylova_n} we used the photometric data given in the final
release of the  2MASS catalog for extended objects
(XSC~\cite{Jarrett:Kopylova_n}). About a half of the galaxies (at the distance of
the UMa system ($z\simeq 0.06$)) discovered spectroscopically in the
SDSS catalog have no measurements in the XSC and we therefore used the data from
the point-source catalog (PSC) for these objects. The magnitudes of bright
galaxies, listed in this catalog, have rather large errors, although
the magnitude corrections for about $14^m$ galaxies (which are usually absent in
the  XSC catalog) are about $0.2^m$. Figure~\ref{point:Kopylova_n} demonstrates,
as an example, the $K(XSC)$--$K(PSC)$ differences for the same galaxies of the
\mbox{A1377 cluster.}

\begin{figure}[tbp]
\setcaptionmargin{0mm}
\onelinecaptionsfalse
\includegraphics[scale=0.38,angle=-90]{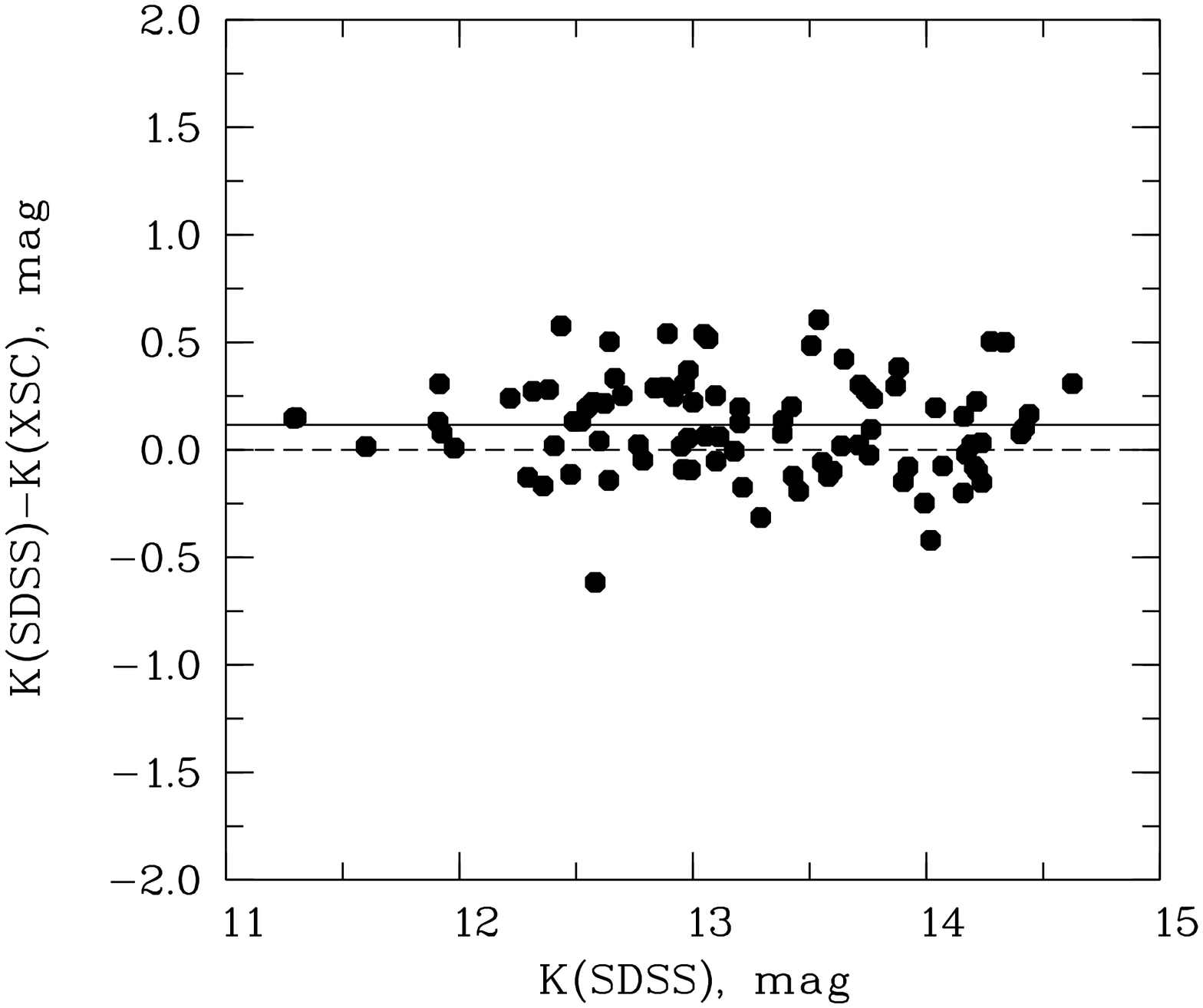}
\captionstyle{normal}
\caption{Comparison of the $K$-band magnitudes computed using the $u-r$ color
index from the SDSS catalog with the 2MASS XCS $\mu_K$~= 20$^m/\square''$
isophotal magnitudes for galaxies of the A1377 cluster. The solid line indicates
the average magnitude difference.}
\label{ksdss:Kopylova_n}
\includegraphics[scale=0.38,angle=-90]{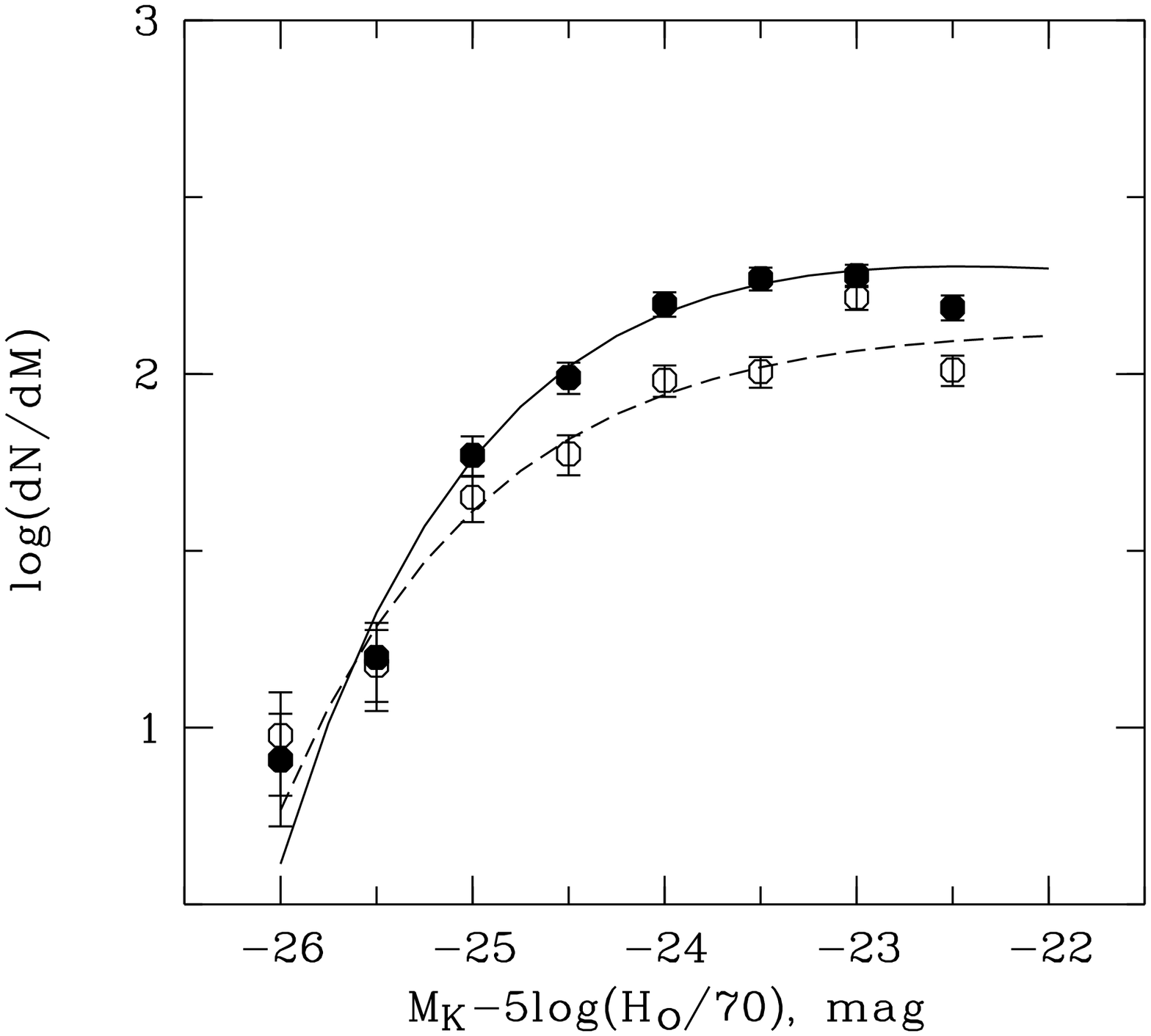}
\captionstyle{normal}
\caption{Combined luminosity functions for UMa galaxies (the filled circles) and
for field galaxies (the open circles). The solid and dashed lines indicate
the corresponding Schechter functions.
}
\label{LFtot:Kopylova_n}
\end{figure}

In this paper to search for the $K$-band magnitudes we use a different method
described by \mbox{M.~Obri\'{c} et al.~\cite{Obric:Kopylova_n}.} They calculate
the $K(SDSS)$ magnitudes for 99000 SDSS (DR1) galaxies based on their
\mbox{$u-r$} color indices. We proceeded as follows: $K(SDSS) = r_{pet}-(r-K)$,
where $r_{pet}$  is the Petrosian $r$-band magnitude of the galaxy
and $r-K$ is given by the formula $r-K =
1.115+0.94(u-r)-0.165(u-r)^2+0.00851(u-r)^3+4.92z-9.1z^2$ \linebreak
(here $z$ is the redshift of the galaxy). We further corrected the  \mbox{$r-K$}
colors by adding the terms \mbox{$0.496-0.154R^z_{50}$} for late-type \mbox{($u-r<2.22$)}
and \mbox{$0.107-0.045R^z_{50}$} for
early-type \mbox{($u-r \geq2.22$)} galaxies, respectively, where $R^z_{50}$ is
the radius of the region emitting $50\%$ of the Petrosian $z$-band flux. We
adjusted Petrosian's magnitudes as described by Graham~\cite{Graham:Kopylova_n}
to transform them into the total galaxy magnitudes by the formula:
$$r_{tot} = r_{pet}-5.1\times10^{-4}\times \exp((R^r_{90}/R^r_{50})^{1.451}),$$
\noindent
where $R^r_{90}$ and $R^r_{50}$ are the radii of the regions containing  $90\%$
and $50\%$ of the Petrosian $r$-band flux, respectively.
Figure~\ref{ksdss:Kopylova_n} represents the difference between the computed
($K(SDSS)$) and integrated ($K(XSC)$) magnitudes obtained from isophotal
magnitudes corresponding to the surface brightness level of $\mu_K = 20^m
/\square''$, minus $0.2^m$ in accordance with the recommendations
of Kochanek et al.~\cite{Kochanek:Kopylova_n}. E.g., the corresponding magnitude
difference for 92 galaxies ($r_{pet}<17.77^m$) of the A1377 cluster is equal to
$(0.12\pm 0.02)^m$, and it is the same for other clusters within the quoted
errors. We applied this correction when computing  $K(SDSS)$. The
average error of the isophotal magnitudes in the extended source catalog is
equal to $0.1^m$ for the clusters  studied.

The completeness of our sample is determined by that of the spectroscopic data
of the SDSS catalog. For objects satisfying the conditions $r_{pet}<17.77^m$ and
\mbox{$\mu_r < 24.5^m/\square''$} (the Petrosian $r$-band magnitude of the
galaxy corrected for Galactic absorption and the average Petrosian surface
brightness corresponding to the effective radius) the completeness of SDSS data
is estimated  at $99\%$ \cite{Strauss:Kopylova_n} and the completeness for
bright galaxies, at $95\%$. We supplemented the radial-velocity measurements for
bright galaxies (one to five objects per cluster), which are unavailable in
SDSS, by adopting them from the  NED database.

The $L_{200,K}$ luminosity of a cluster is equal to the sum of the $K$-band
luminosities of its member galaxies located within $R_{200}$ down to a
fixed limiting absolute magnitude. We set this limit, like Lin et
al.~\cite{Lin:Kopylova_n}, equal to  $-21^m$. We first transformed the
observed galaxy magnitudes into the corresponding absolute magnitudes by the
formula: \linebreak
\noindent
$M_K = K-25-5 \log_{10}(D_{l}/1$\,Mpc$)-A_K-K(z),$
\noindent
where $D_l$ is the distance to the galaxy used to compute its luminosity; $A_K$
is the Galactic extinction and
\mbox{$K(z)$ = $-6\log(1+z)$}, the K-correction computed  \mbox{in accordance
with
\cite{Kochanek:Kopylova_n}.}
We applied no evolutionary corrections because of the small redshift
interval (from 0.045 to 0.075) spanned by the objects in our sample. Galactic
extinction, which we adopt from NED, is less than $0.01^m$ for our galaxies.
The 2MASS (XSC) is not a deep survey (its $K$-band photometric limit for the
completeness level of 90\% and above is equal to $13.5^m$ in $K$-band
\cite{Jarrett:Kopylova_n}). We supplemented this catalog with galaxies from SDSS catalog
with the Petrosian magnitudes $r_{pet}<17.77^m$. Given that the
$r-K$ color index is, on the average, equal to $2.8^m$ for early-type galaxies,
which make up most of galaxies within $R_{200}$, the limiting $K$-band magnitude of
our sample of galaxies with individual $K$-band magnitude estimates must be of
about $15^m$.

The computation of the total luminosity of a cluster of galaxies using the
luminosity function (LF) inside the selected radius is a two-stage
procedure~\cite{Ramella:Kopylova_n}: the LF should first be normalized to the observed
number of galaxies and then extrapolated into the domain of faint magnitudes
down to the adopted limit. This is generally done either using the parameters of
the Schechter function (the characteristic magnitude $M_K^*$ and the slope
$\alpha$) for field galaxies or using the corresponding parameters determined
for the composite LF of the sample studied. To estimate how the
computed luminosities of clusters differ in these two cases, we first used the
parameters of the Schechter function for field galaxies as inferred by \mbox{Kochanek
et al.~\cite{Kochanek:Kopylova_n}}
($M_K^*$, $\alpha$) = ($-24.14^m$, -1.09), which are  commonly employed by other
authors. We computed the normalization parameter for each cluster:
\begin{equation}
\nonumber
\phi^*=N_{obs}/\int^{+\infty}_{L_{K,lim}/L_K^*}t^{\alpha} e^{-t} dt,
\end{equation}
where $t=L/L_K^*$.
We then summed up the luminosities of galaxies down to the limiting magnitude of
$15^m$ and used the normalization quantity obtained to extrapolate the LF downward
to the limiting magnitude of  $-21^m$, which corresponds
to the luminosity $L_{K,min}$:
\begin{equation}
\nonumber
L_{K}=\sum_{i=1}^{N_{obs}}L_{{K},i}+\phi^*L^*_{K}\int^{L_{K,lim}/L_K^*}
_{L_{K,min}/L_K^*}t^{\alpha+1} e^{-t} dt.
\end{equation}
This procedure increases the luminosity of the cluster by  5\% on the average.

In the second variant of the computation of the total luminosities of clusters
we searched for the parameters of the Schechter function for composite
luminosity function of our sample. To perform this, we counted galaxies in
each cluster within  $0.5^m$-wide magnitude intervals. We then constructed the
composite luminosity functions for the virialized regions of UMa clusters and
for the clusters located in less densely populated immediate environments using
the method described by Colless~\cite{Colless:Kopylova_n} and approximated
the resulting composite LFs via nonlinear least squares method by the Schechter
functions~\cite{Schechter:Kopylova_n}
with the parameters $M^*_K=-24.33^m \pm0.04$, $\alpha=-0.82\pm0.02$ and
$M^*_K=-24.20^m \pm0.13$, $\alpha =-0.92\pm0.09$ for UMa clusters
and for the clusters from the UMa neighborhood, respectively (we chose our last
data point in the domain of faint magnitudes to be $-23.0^m$ in order to avoid
data incompleteness for more distant clusters). We compute the errors as
$\sqrt{dN}$, where $dN$ is the number of galaxies in the  $dM$ interval. The
exclusion of the highest-luminosity data point from the fit has no significant
effect on the parameters above. Figure~\ref{LFtot:Kopylova_n} shows both
luminosity functions. The parameters of two functions are almost the same
and we therefore constructed the composite luminosity function for all clusters
and used the Schechter function parameters inferred for the combined sample
\mbox{($M^*_K=-24.29^m \pm0.05$,} \mbox{$\alpha =-0.85\pm0.03$)} to compute
the total cluster luminosities listed in  Table~\ref{data2:Kopylova_n}. The
total luminosities of clusters of galaxies, calculated using this way, on the average are
higher than the corresponding luminosities computed in the first variant by
$3\%$, implying an increase of the mass-to-luminosity ratio in the same
proportion. The parameters of the luminosity functions for individual clusters
differ by less than  $3\sigma$ from the values above.
\begin{table}[tbp]
\setcaptionmargin{0mm}
\onelinecaptionstrue
\captionstyle{flushleft}
\caption{Properties of clusters in the near IR}
\label{data2:Kopylova_n}
\medskip
\begin{tabular}{l|c|c|c|l} \hline
Cluster& $L_{K,200}$& $M/L_K$& $N(-21^m)$& $f_E$($N_E$)\\
     & $10^{12}~L_{\odot}$& $M_{\odot}/L_{\odot}$&  &\\
\hline
A1270   & $5.12\pm0.16$ & $55\pm18$&   92&   0.63(22)  \\
A1291A  & $1.82\pm0.10$ & $55\pm30$&   41&   0.85(11)  \\
A1291B  & $1.78\pm0.12$ & $158\pm80$&  50&   0.69(9)   \\
A1318   & $2.90\pm0.17$ & $36\pm17$&   53&   0.78(11)  \\
A1377   & $6.52\pm0.15$ & $66\pm21$&   103&  0.65(26)  \\
A1383   & $3.76\pm0.13$ & $45\pm16$&   72&   0.67(20)  \\
A1436   & $6.74\pm0.16$ & $79\pm23$&   130&  0.82(37)  \\
Anon1   & $4.57\pm0.14$ & $83\pm28$&   84&   0.71(27)  \\
Anon2   & $1.96\pm0.16$ & $14\pm8$&    23&   0.91(10)  \\
Anon3   & $1.88\pm0.16$ & $47\pm23$&   35&   0.78(11)  \\
Anon4   & $2.36\pm0.14$ & $44\pm20$&   33&   0.50(8)   \\
Sh166   & $1.35\pm0.09$ & $44\pm23$&   36&   0.70(7)   \\
A1003   & $2.82\pm0.16$ & $106\pm55$&  41&   0.47(8)   \\
A1169   & $4.25\pm0.11$ & $78\pm27$&   89&   0.68(19)  \\
A1279   & $0.64\pm0.14$ & $17\pm16$&   8&    0.25(1)   \\
A1452   & $2.08\pm0.09$ & $109\pm64$&  29&   0.67(8)   \\
A1461   & $0.82\pm0.10$ & $66\pm49$&   15&   0.75(3)   \\
A1507   & $2.63\pm0.14$ & $52\pm21$&   50&   0.71(12)  \\
A1534   & $2.29\pm0.14$ & $21\pm12$&   29&   0.69(9)   \\
RXCJ1010& $2.00\pm0.16$ & $62\pm23$&   39&   0.64(7)   \\
RXJ1033 & $2.69\pm0.18$ & $45\pm18$&   52&   0.56(9)   \\
RXCJ1053A&$3.94\pm0.11$ & $55\pm21$&   80&   0.68(19)  \\
RXCJ1053B&$2.59\pm0.08$ & $47\pm35$&   46&   0.58(11)  \\
RXCJ1122& $0.79\pm0.13$ & $24\pm18$&   16&   0.80(4)  \\
\hline
\end{tabular}
\end{table}

Unlike the other authors, we found that parameter $M^*_K$ in virialized
regions of galaxies clusters located in the UMa region is $0.15^m$ lower than the same
for field galaxies \cite{Kochanek:Kopylova_n}. In the virialized regions of nine
nearby rich  (495\,km/s  < $\sigma$ < 1042\,km/s ) and \mbox{X-ray} luminous  clusters
of galaxies the $M^*_K$ parameter is lower by $0.62^m$ \cite{Rines:Kopylova_n}
and the fraction of faint galaxies is much higher ($\alpha$=--1.35).
\mbox{Rines et al.~\cite{Rines:Kopylova_n}} used the total extrapolated
magnitudes.

\section{PROPERTIES OF EARLY-TYPE GALAXIES IN THE VIRIALIZED REGIONS OF CLUSTERS
OF GALAXIES}
\begin{figure}[tbp]
\setcaptionmargin{0mm}
\onelinecaptionsfalse
\includegraphics[scale=0.31,angle=-90]{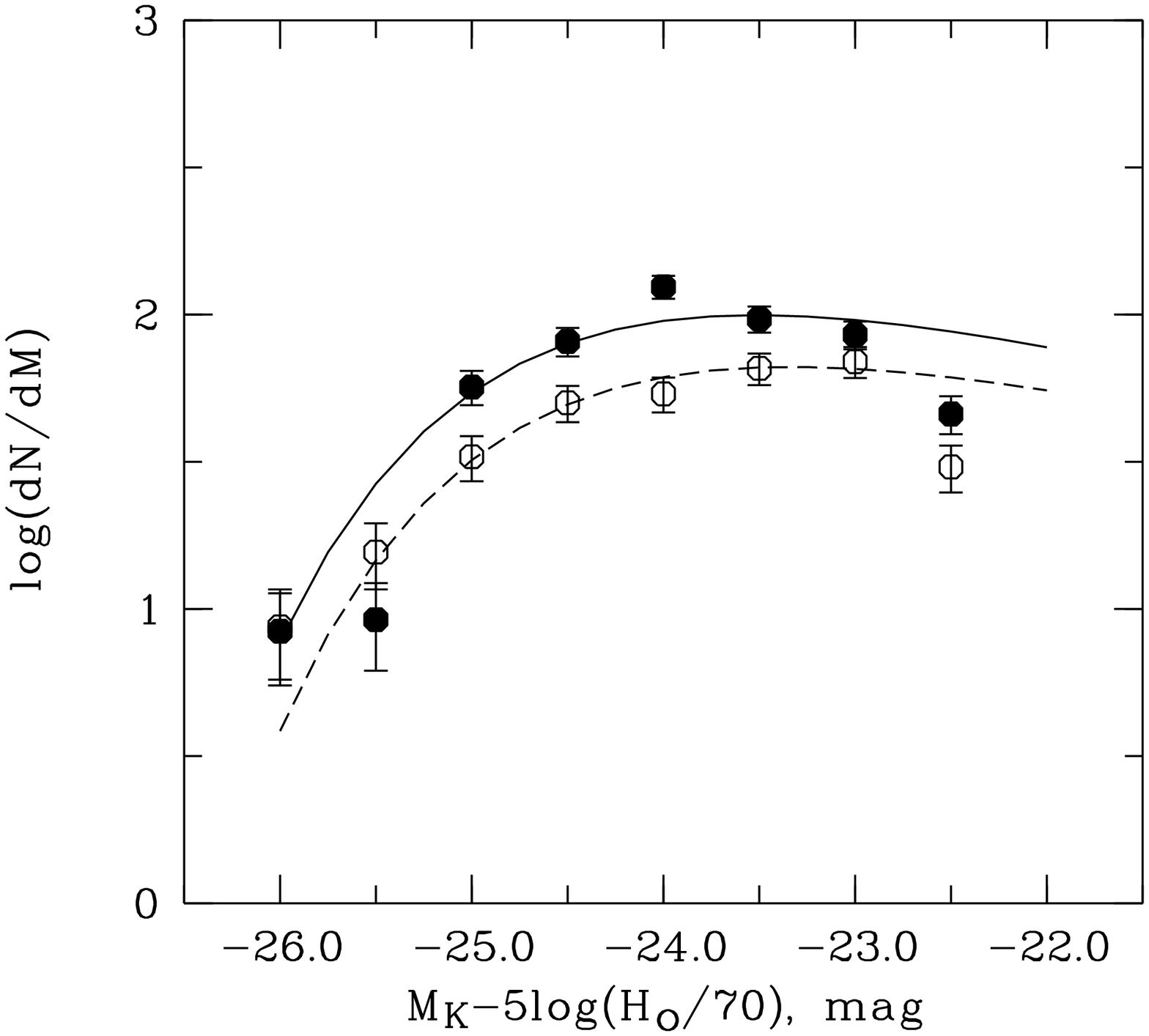}
\captionstyle{normal}
\caption{Composite luminosity functions for early-type galaxies. Same
designations as in Fig.~\ref{LFtot:Kopylova_n}.}
\label{LufE:Kopylova_n}
\includegraphics[scale=0.31,angle=-90]{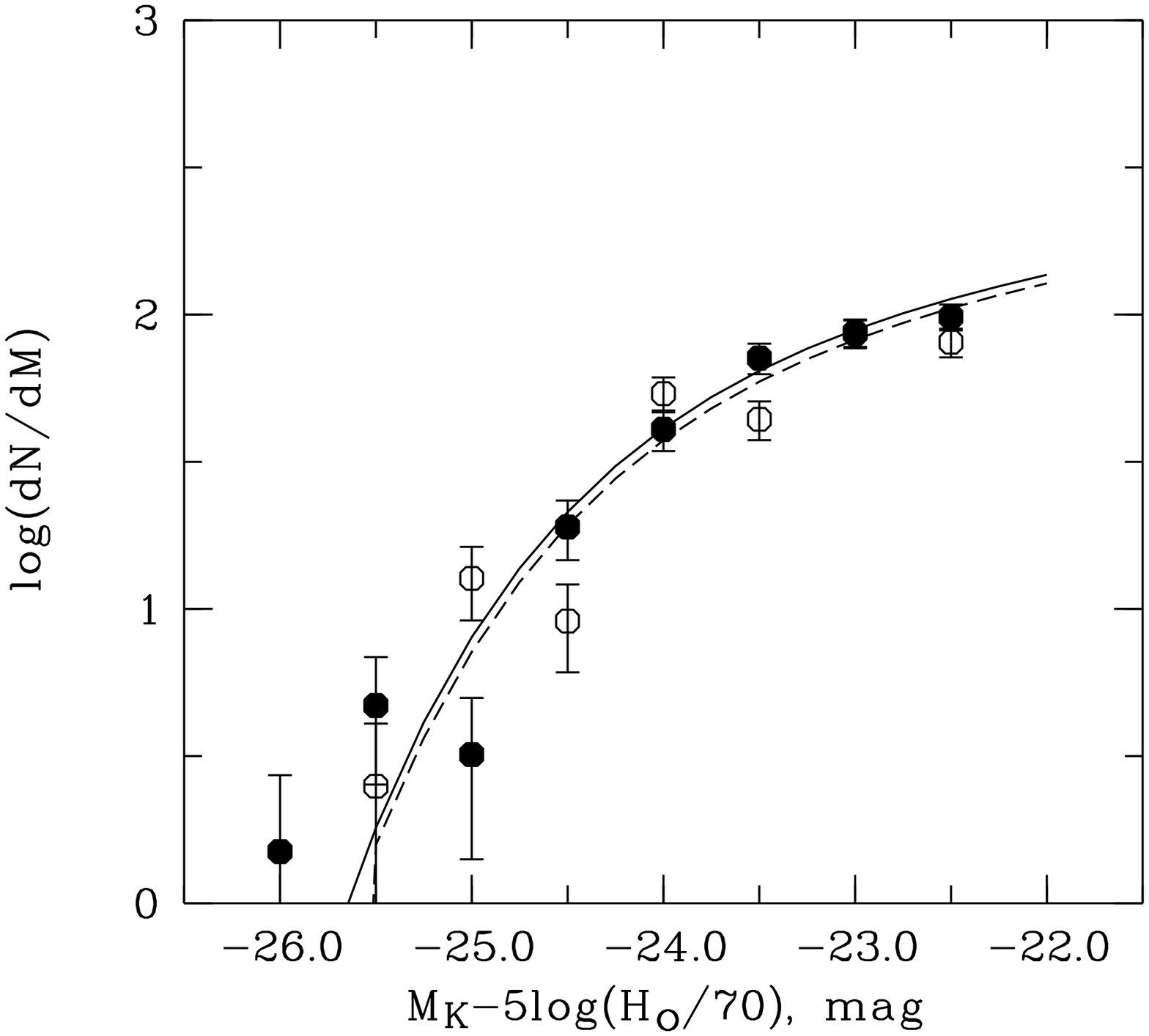}
\captionstyle{normal}
\caption{Composite luminosity functions for late-type galaxies.
Same designations as in Fig.~\ref{LFtot:Kopylova_n}.}
\label{LufS:Kopylova_n}
\includegraphics[scale=0.31,angle=-90]{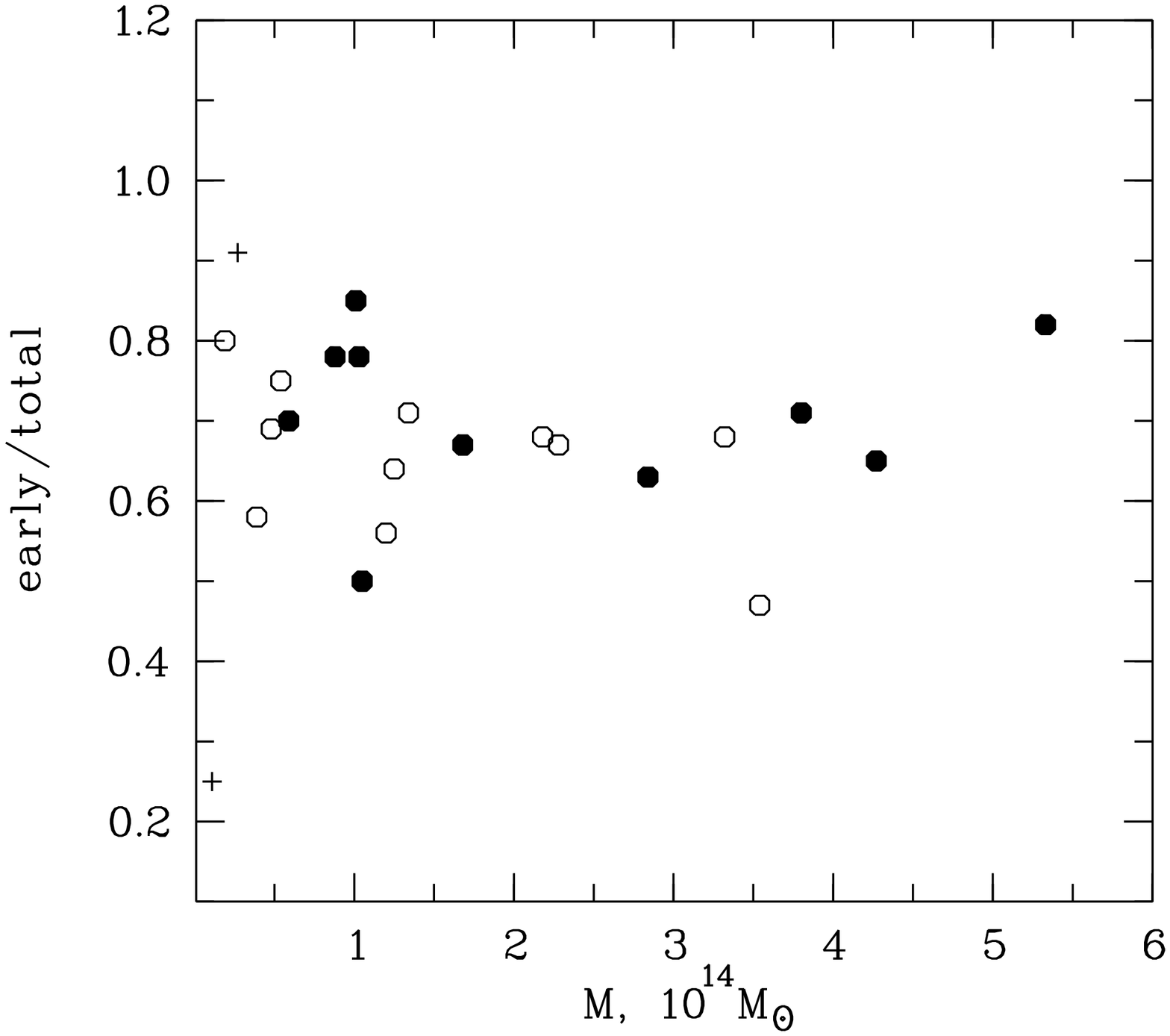}
\captionstyle{normal}
\caption{The fraction of early-type galaxies among the galaxies brighter than
$M^*_K+1$ as a function of the cluster mass inside $R_{200}$.
The filled and open circles correspond to the clusters located in the UMa and in
its neighborhood, respectively. The plus signs indicate the clusters with the
smallest (A1279) and greatest (Anon2) fractions of early-type galaxies.}
\label{ratE:Kopylova_n}
\end{figure}

The fraction of early-type galaxies in clusters is known to increase with
increasing local galaxy density \cite{Dressler:Kopylova_n,Goto:Kopylova_n},
and so does the luminosity of their spheroidal \mbox{components
\cite{Dressler:Kopylova_n}.} In this paper we analyze the properties of
early-type galaxies located in the UMa supercluster and in its
immediate vicinity. We used the following criteria to select early-type galaxies
(in the $r$-band filter of SDSS): $fracDeV \geq 0.8$ (this parameter
characterizes the contribution of the de Vaucouleurs profile to the
surface-brightness profile of the galaxy); $R^r_{90}/R^r_{50}\geq 2.6$ (the
concentration index is equal to the ratio of the radii containing
$90\%$ and $50\%$ of the Petrosian flux). We constructed the composite
$K$-band luminosity functions separately for early- and late-type galaxies
(Figs.~\ref{LufE:Kopylova_n} and \ref{LufS:Kopylova_n}) in the virialized
regions of clusters of galaxies. We derived the following parameters for the
Schechter function:
\begin{itemize}
\item[(a)]
\mbox{$M_K^*=-24.54^m \pm0.18$,} \mbox{$\alpha =-0.61\pm0.09$}  for early-type
galaxies of the UMa system
and  \linebreak
\mbox{$M_K^*=-24.48^m \pm0.07$,} \mbox{$\alpha =-0.65\pm0.04$} for galaxies in
the UMa neighborhood;
\item [(b)]
\mbox{$M_K^*=-24.06^m \pm0.06$}, \mbox{$\alpha =-1.22\pm0.06$}
for late-type galaxies of the UMa system
and  \linebreak
$M_K^*=-24.05^m \pm0.18$, \mbox{$\alpha =-1.23\pm0.18$}
for galaxies in the UMa neighborhood.
\end{itemize}
The number of early-type galaxies in 10 UMa clusters is  $40\%$ greater than the same in
11 clusters in the  nearest neighborhood of the supercluster, and the number of
late-type galaxies is  $15\%$ greater down to the adopted limiting magnitude of
($-21^m$). We compared the $M^*_K$ for early-type galaxies
in the UMa region~\mbox{($-24.54^m$, $-24.48^m$)} with the similar values for
field galaxies~\cite{Kochanek:Kopylova_n} ($-24.28^m$). We performed a similar
comparison of the $M^*_K$ magnitudes for late-type galaxies in the UMa
neighborhood ($-24.06^m$, $-24.05^m$) and for field galaxies
\mbox{($-23.73^m$)} from the same paper. We found that the $M^*_K$  magnitudes
in the virialized regions of clusters of galaxies to be $0.23^m$ and $0.32^m$
lower for early- and late-type galaxies, respectively, located in the UMa region
than the corresponding magnitudes for field galaxies. Moreover, the number
of early-type galaxies in the clusters located in the region studied decreases
($\alpha >-1$) and that of late-type galaxies increases ($\alpha <-1$) with
decreasing galaxy luminosity.

We calculate the fraction of early-type galaxies  (i.e., the ratio of
number of early-type galaxies to the total number of galaxies) inside
$R_{200}$ down to a fixed limiting magnitude corresponding to  $M_K^*+1$,
where $M_K^*$ is equal to the characteristic magnitude of the cluster sample
($-24.29^m$). We found that, on the average, the fraction of bright early-type
galaxies in the UMa system is equal to $0.71 \pm 0.03$ (without A1291B and
Anon2), and the corresponding fraction for the neighborhood of the UMa
supercluster is   $0.66\pm 0.03$ (without A1279). Figure~\ref{ratE:Kopylova_n}
presents the fraction of bright early-type galaxies plotted as a function of the
cluster mass. It is evident from the figure that the fraction of
early-type galaxies remains almost unchanged with the cluster mass lying in the
interval \mbox{$10^{13}$~$M_{\odot} < M < 5.0\times
10^{14}$~$M_{\odot}$}. A similar result for clusters of galaxies was obtained,
e.g.,  \mbox{by Balogh et al. \cite{Balogh:Kopylova_n} and
Tanaka et al.~\cite{Tanaka:Kopylova_n}.} At the same time, Martinez et
al.~\cite{Martinez:Kopylova_n} and \linebreak \mbox{Weinmann et
al.~\cite{Weinmann:Kopylova_n}} found that the fraction of such galaxies in
groups decreases with the decreasing mass of the group. A1279 and Anon2 with
their extremely large early-type galaxy populations stand out among the clusters
of galaxies studied in this paper and are marked by \mbox{plus signs in
Fig.~\ref{ratE:Kopylova_n}.}

We found the average parameters of bright early-type galaxies (down to the
fixed magnitude of  $M_K^*+1$), the  $K$-band absolute magnitude, the $g-r$ and
$u-r$ color indices, the effective bulge radius in the $r$ band, the axial
ratio $b/a$, the contribution of the de Vaucouleurs bulge to the
surface-brightness profile and to the concentration index, which is equal to the
ratio of the galactocentric radii containing $50\%$ and $90\%$ of the Petrosian
flux. Table~\ref{data3:Kopylova_n} lists the data obtained for the UMa
supercluster, its immediate vicinity, and three samples of galaxies. Cluster
subsamples are selected  from the list of clusters studied, which subdivided by
mass:
\begin{itemize}
\item[(I)] M>$3\times10^{14}$ $M_{\odot};$
\item[(II)]$1\times10^{14}$ $M_{\odot}$<M<$3\times10^{14}$ $M_{\odot};$
\item[(III)]M<$1\times10^{14}$ $M_{\odot}.$
\end{itemize}
Sample I consists of the most massive clusters: A1003, A1169, A1377, A1436, and
Anon1; sample~II consists of A1270, A1291A, A1318, A1383, A1452, A1507, Anon4,
RXCJ1010, RXJ1033, RXCJ1053A, and RXCJ1053B, and sample~III consists of A1461,
A1534, Anon3, Sh166, and RXCJ1122. An analysis of the data listed in Table\,3
leads us to conclude that early-type (i.e., bulge dominated) galaxies have
nearly the same properties in all samples. There is a weak tendency for
these galaxies to be brighter in the IR on the one hand, in field clusters in
the vicinity of UMa, and, on the other hand, in less massive clusters (i.e., in
those with lower velocity dispersion).
\begin{table*}[tbp]
\setcaptionmargin{0mm}
\onelinecaptionstrue
\captionstyle{flushleft}
\caption{Average parameters  for early-type galaxies ($M_K<-23.29^m$) and their
dispersions}
\label{data3:Kopylova_n}
\medskip
\begin{tabular}{c|r|r|c|c|c|c|c|c|c} \hline
      & $N_c$& $N_g$& $M_K$, & $g-r$, & $u-r$, & $r_e$,& $b/a$& $fracDeV$ &
$R^r_{90}/R^r_{50}$\\
      &   &    & mag & mag   &  mag    &  kpc  &      &       &
\\
\hline
UMa    & 10& 180& $-24.19(0.59)$& $0.85(0.04)$& $2.71(0.16)$& $4.12(1.62)$&
$0.69(0.18)$& $0.97(0.06)$& $3.05(0.24)$\\
field  & 11& 109& $-24.32(0.68)$& $0.86(0.04)$& $2.73(0.12)$& $4.41(2.17)$&
$0.69(0.19)$& $0.97(0.05)$& $3.02(0.22)$\\
I     &  5& 117& $-24.20(0.60)$& $0.85(0.04)$& $2.70(0.14)$& $4.41(1.64)$&
$0.70(0.18)$& $0.97(0.06)$& $3.02(0.23)$\\
II    & 11& 138& $-24.25(0.65)$& $0.86(0.04)$& $2.72(0.15)$& $4.06(1.94)$&
$0.67(0.18)$& $0.97(0.05)$& $3.05(0.24)$\\
III   &  5&  34& $-24.32(0.64)$& $0.86(0.04)$& $2.75(0.14)$& $4.60(1.95)$&
$0.72(0.20)$& $0.97(0.05)$& $3.09(0.23)$\\
\hline
\end{tabular}
\end{table*}

\section{THE RELATION BETWEEN THE TOTAL $K$-BAND LUMINOSITIES AND MASSES OF
CLUSTERS}

\begin{figure}[tbp]
\setcaptionmargin{0mm}
\onelinecaptionsfalse
\includegraphics[scale=0.29,angle=-90]{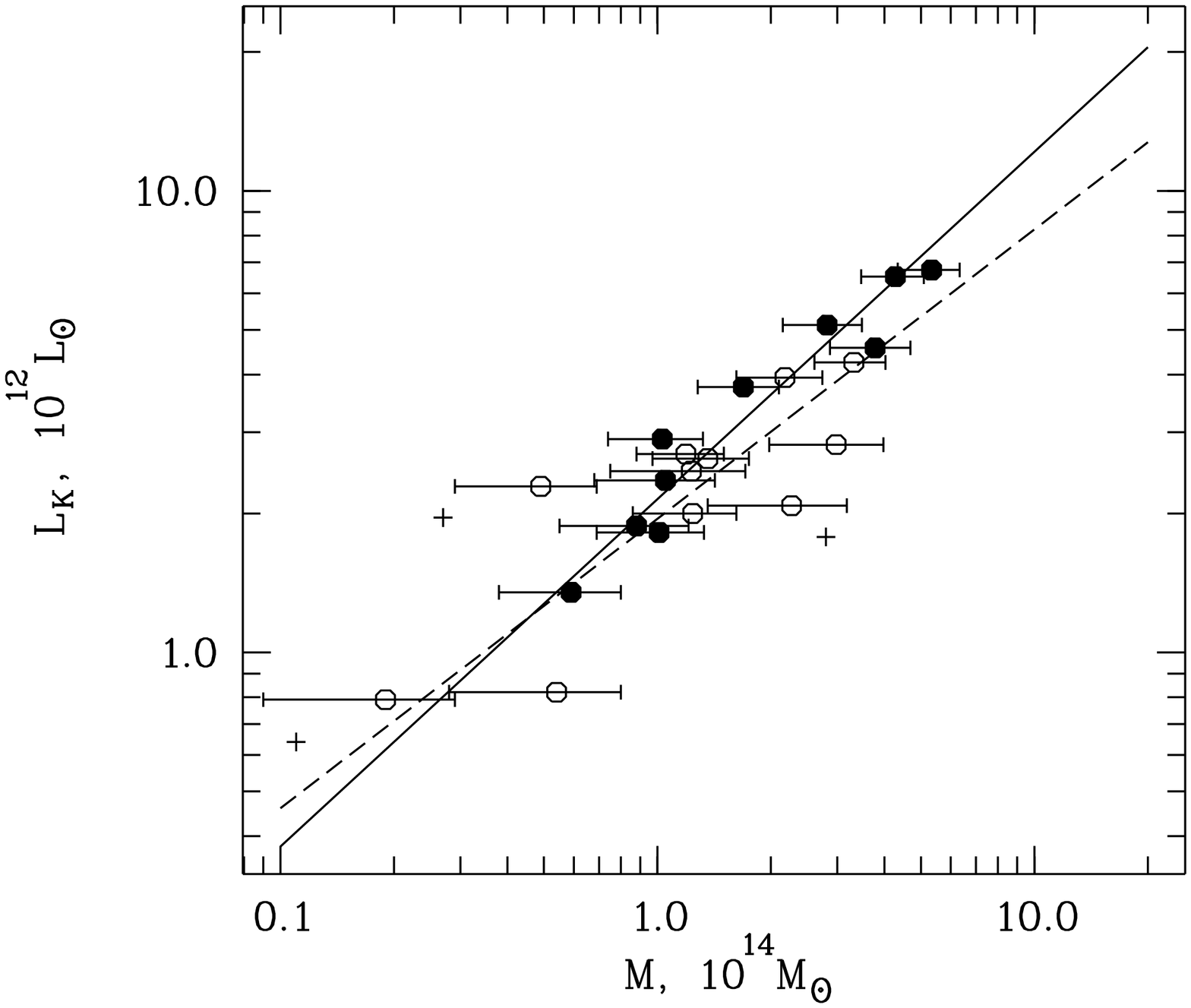}
\captionstyle{normal}
\caption{Total luminosity of galaxies ($M^*_K<-21^m$) as a function of the
cluster mass inside $R_{200}$. The filled and open circles
correspond to UMa and field clusters, respectively. The plus signs
indicate the clusters A1279, Anon2, and A1291B, which were not used to derive
the regression relations. These relations for UMa and field clusters are shown by the solid and
dashed lines, respectively.}
\label{LM:Kopylova_n}
\includegraphics[scale=0.29,angle=-90]{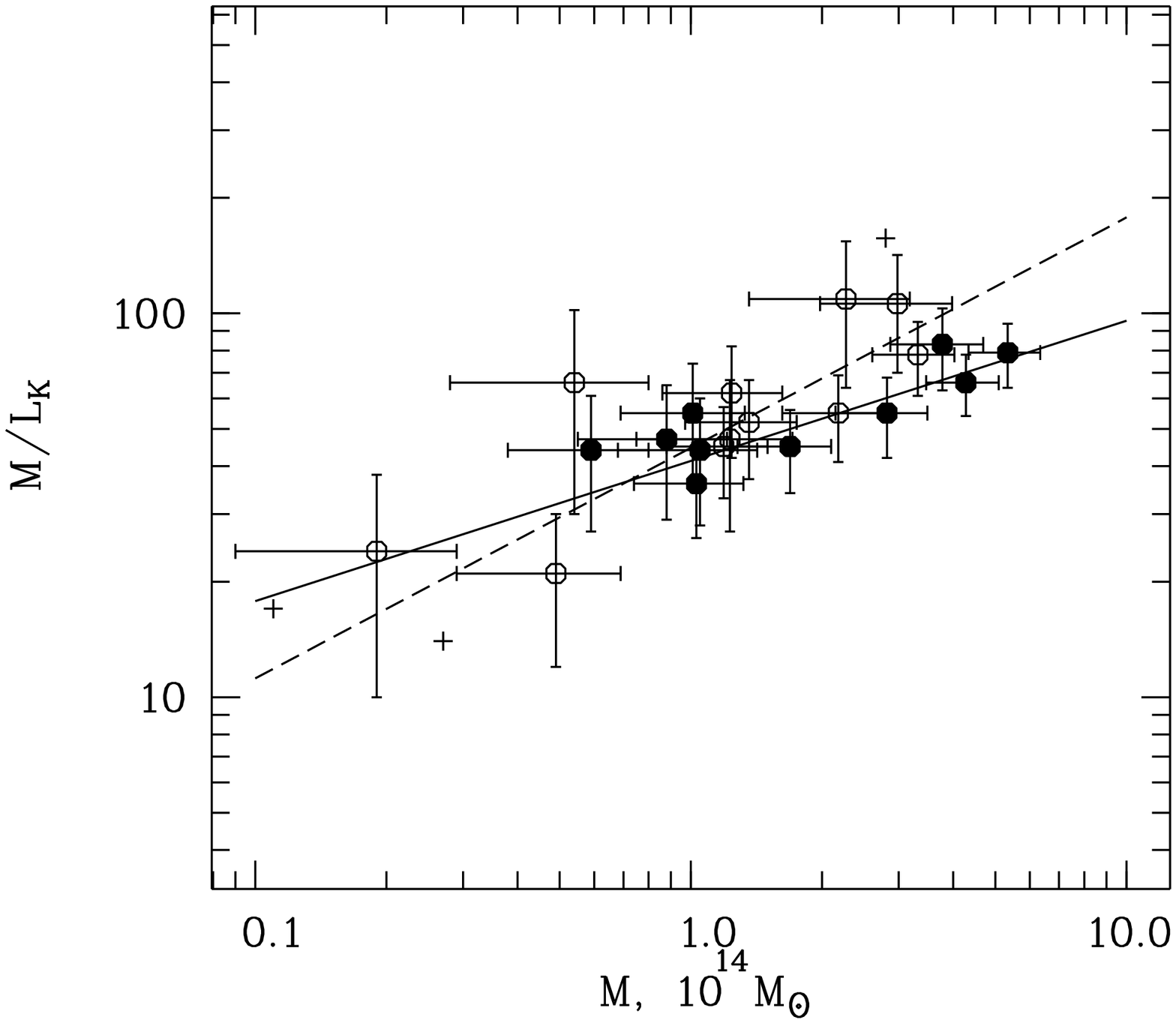}
\captionstyle{normal}
\caption{The mass-to-light ratio ($M^*_K<-21^m$) as a function of the
cluster mass inside $R_{200}$. Designations are the same as in
Fig.~\ref{LM:Kopylova_n}.}
\label{MLM:Kopylova_n}
\includegraphics[scale=0.29,angle=-90]{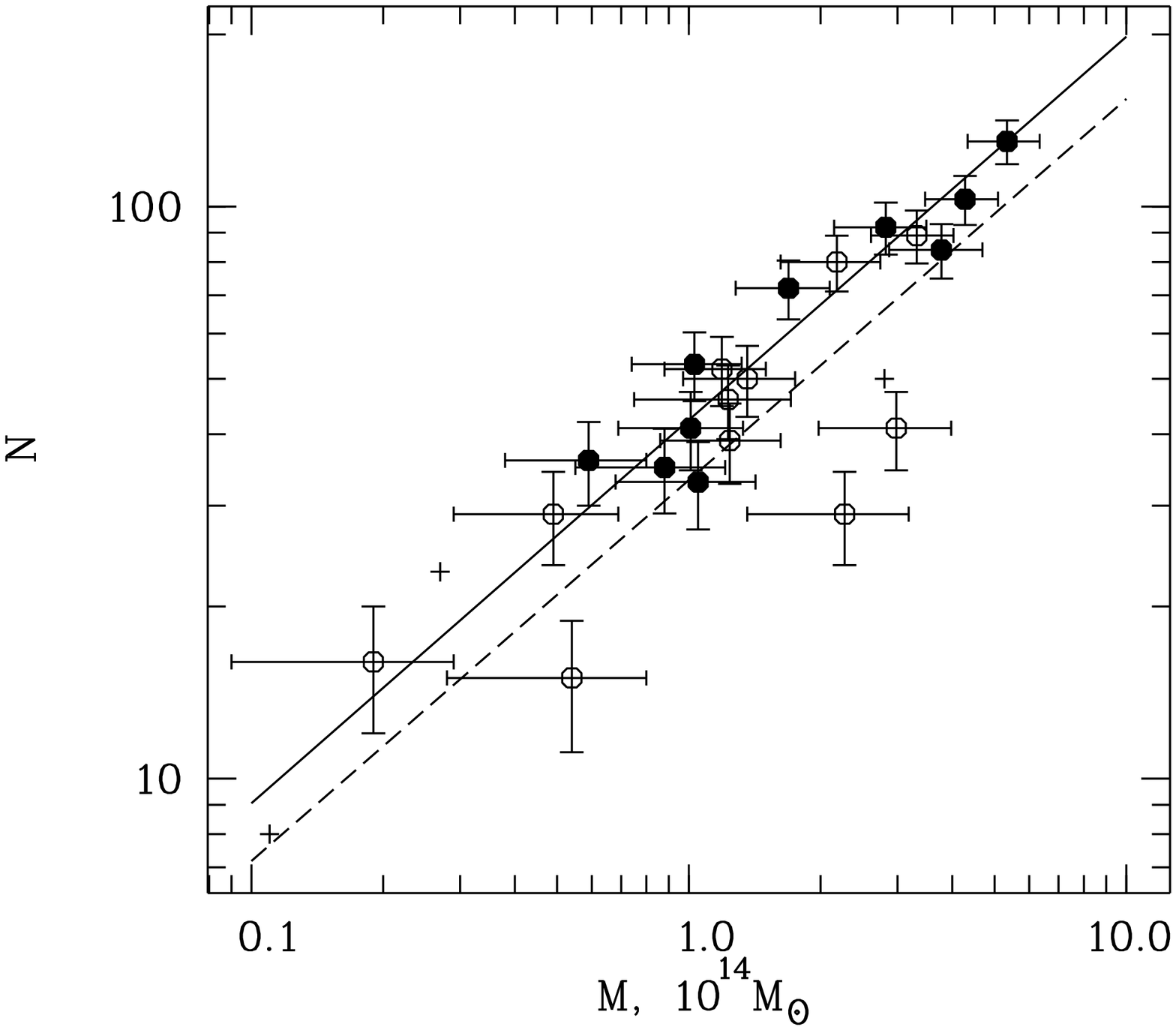}
\captionstyle{normal}
\caption{The number of galaxies in the cluster ($M^*_K<-21^m$) as a function of
the cluster mass inside $R_{200}$. Designations are the same as in
Fig.~\ref{LM:Kopylova_n}.}
\label{NM:Kopylova_n}
\end{figure}

The measurements of the $K$-band luminosities of clusters and groups of galaxies
(mostly of the virialized regions)
\cite{Kop2:Kopylova_n,Lin:Kopylova_n,Ramella:Kopylova_n,Rines:Kopylova_n} showed
that the mass-to-luminosity ratio and the luminosities of clusters increase with
increasing mass of the system (the mass of the dark halo). In this paper
we find the relations between these parameters for the particular region: the
UMa supercluster and its nearest neighborhood. As we already pointed out above,
we find the total luminosities of clusters of galaxies using the Schechter
parameters for the composite luminosity function ($-24.29^m$, --0.85). The
relations between the cluster mass and  $K$-band luminosity for the UMa
supercluster and field clusters in its immediate vicinity have the following
form (in logarithmic form):
$$\log L_{K,200} = 0.75(\pm0.10) \log M_{200}+ 1.80(\pm1.22),$$
$$\log L_{K,200} = 0.63(\pm0.18) \log M_{200}+ 3.50(\pm2.36).$$
We present these relations in Fig.~\ref{LM:Kopylova_n}. The relations obtained are
averages of direct and inverse regression derived by treating luminosity and
mass, respectively, as independent variables. The rms scatter is equal to 0.08
and 0.16 for the relations derived for UMa system and field clusters,
respectively. The scatter in the supercluster is about twice smaller than in the
field. Note that the slope of the relation between  $M_{200}$ and $L_{K,200}$
for clusters of galaxies in the UMa coincides within the quoted errors with the
corresponding slope for field clusters. The luminosities of clusters of galaxies
are computed in projection, i.e., within cylinders, and deprojection may
decrease them by about  $20\%$ (depending on the location of the cluster). The
luminosity corrections should be minimal, because the filamentary structures in
the supercluster are aligned parallel to the sky plane 
\mbox{(see [6, Fig. 2]).}
A comparison  shows that despite of use of
different techniques our results agree with those obtained by other authors:
\mbox{$L_{K}~ \propto~M^{0.72\pm0.04}$~\cite{Lin:Kopylova_n}} and
\mbox{$L_{K}~\propto~M^{0.64\pm0.06}$~\cite{Ramella:Kopylova_n}.}
The correlation between the masses and luminosities of clusters of galaxies
allows the masses of individual clusters to be estimated from their
\mbox{$K$-band luminosities.} The scatter of this relation is of special
interest. This is believed to be most likely due to the deviation of the
dynamical state of the cluster from virial equilibrium (see,
e.g.,~\cite{Ramella:Kopylova_n}). In our sample the clusters those with the greatest
deviations from the mean value are Anon2 and A1291B from UMa and A1279, which a
very poor cluster located in the UMa neighborhood. We did not use them for
computing our regression models. The A1003, A1452, and A1461 clusters
in the UMa neighborhood also deviate from the average relation due to their
higher mass (dispersion), which appears to correspond to non-equilibrium state,
and the A1534 cluster---due to its lower mass. The relation between the mass and the
total luminosity can actually be used to identify clusters of galaxies that are
in a peculiar state, with either too high or too low velocity (mass) dispersion
combined with rather high luminosity. The rather small scatter that we obtained for
the clusters belonging to the UMa supercluster (which is the densest part of the
system of clusters) and which exceeds the scatter obtained in other similar
studies based on various samples of clusters, which do not form a supercluster,
may be indicative of the synchronous nature of the evolution of clusters
within a single system (UMa). We use the same procedures as in case of the
relation between  $L_{K,200}$ and $M_{200}$ to derive a formula to connect the
logarithms of $M_{200}/L_{K,200}$ and $M_{200}$ for the UMa supercluster and for
the clusters located in the UMa field. The resulting relations have the
following form:
{\small $$\log M_{200}/L_{K,200} = 0.36(\pm0.15)\log M_{200}-3.49(\pm0.96),$$
$$\log M_{200}/L_{K,200} = 0.60(\pm0.18)\log M_{200}-6.75(\pm1.01)$$}
and are shown in Fig.~\ref{MLM:Kopylova_n}. The scatter is the same as that of
the relations between the mass and luminosity. It is interesting that the slope of
these relations is higher by about one third than the slopes of the relations
$M/L\propto M^{0.25}$ and \mbox{$M/L\propto M^{0.37}$} derived from the
relations between $L_{K,200}$ and $M_{200}$ given above. Moreover, the slope of
the dependence of $M/L_K$ on $M$ for clusters in the UMa neighborhood is
somewhat steeper than that for UMa clusters and is equal to
the corresponding slope for groups of galaxies given below. Various authors
obtained the following results for other samples of clusters of galaxies:
\mbox{{\small $M/L_{K}\propto M^{0.26\pm0.04}$}~\cite{Lin:Kopylova_n}} and
\mbox{{\small $M/L_{K}\propto M^{0.31\pm0.09}$}~\cite{Rines:Kopylova_n},}  but
{\small $M/L_{K}\propto M^{0.56\pm0.05}$}~\cite{Ramella:Kopylova_n} for groups
of galaxies. The average  $M/L$ ratio inside  $R_{200}$ for the UMa supercluster
is equal to \mbox{$55\pm5$~$M_{\odot}/L_{\odot}$,} and the corresponding ratio
for field clusters in the UMa neighborhood is $60\pm8$~$M_{\odot}/L_{\odot}$.
Note for comparison, that the average $M/L$ ratio inside $R_{200}$ for
rich clusters of galaxies (with the  masses estimated using the method of
caustics in spheres and the luminosities computed inside the cylinders, like in
this paper) from \cite{Rines:Kopylova_n} is $49\pm5$ $M_{\odot}/L_{\odot}$,
which agrees with our result within the quoted errors. The existence of the
$L$--$M$ and $M/L$--$M$ relations indicates that the formation of stars or
galaxies in clusters and groups are regular processes, however, the efficiency
of star formation inside virialized regions of the clusters may be a decreasing
function of the cluster mass \mbox{(see, e.g., \cite{Lin:Kopylova_n}),} as
evidenced by the smaller-than-unity slope in Fig.~\ref{LM:Kopylova_n}. Note that
the unaccounted contribution of the radiation from intragalactic and
intracluster stars is an increasing function of mass and may amount to 5--50$\%$
(\mbox{see, e.g.,\cite{Rines:Kopylova_n}).}

\subsection{Number of Galaxies in Clusters}
Clusters of galaxies are massive dark halos where baryonic matter in
the form of gas and galaxies is located in the places of highest
concentration. The number of galaxies down to a certain limiting magnitude
located in the cluster inside a certain radius  (the halo occupation number)
is the main parameter, which can be used to compare the results of model
computations with observations. To compute the number of galaxies in a
cluster located within $R_{200}$, we adopted the Schechter function
parameters for the composite luminosity function ($-24.29^m$, --0.85),
extrapolated this function down to a certain limiting magnitude
equal to $-21^m$. The normalization factor is equal to the number of observed
galaxies down to the limiting magnitude of  $15^m$. We derived the relations
between the logarithms of  $N_{200}$ and $M_{200}$ for the UMa supercluster and
for clusters in the UMa field using the same procedures as those employed to
derive the previous relations. Figure~\ref{NM:Kopylova_n} presents the resulting
relations, which have the following form:
$$\log N_{200} = 0.67(\pm0.10) \log M_{200}-7.75(\pm0.65),$$
$$\log N_{200} = 0.67(\pm0.21) \log M_{200}-7.80(\pm1.21).$$
The scatter is the same as in case of the previous relations---it is
twice smaller for UMa clusters compared to the clusters located in the
immediate neighborhood of the supercluster. The slope of the relation between
$N$ and $M$ is the same for both subsamples in what appears to be
consistent with the results of some studies while differing slightly
from those of other investigations (possibly, due to the methods used to
estimate the errors). For example, the following relations were derived for a
sample of rich clusters:
$N \propto M^{0.70\pm0.09}$~\cite{Rines:Kopylova_n} for $R_{200}$ and
$N \propto M^{0.84\pm0.04}$~\cite{Lin:Kopylova_n} for $R_{500}.$
For groups of galaxies within $R_{200}$ the relation
$N \propto M^{0.56\pm0.05}$~\cite{Ramella:Kopylova_n} was derived. From the
theoretical viewpoint the slope must be smaller than unity (it must lie in
the 0.56--0.74 interval), i.e., the efficiency of the formation of galaxies
must be lower and/or the processes of galaxy disruption must be less
efficient in more massive halos (see, e.g.,  \mbox{\cite{Rines:Kopylova_n}}
and references therein). We list the inferred parameters of clusters of
galaxies in Table~\ref{data2:Kopylova_n}. The number  $N(-21)$ of galaxies
in the cluster is equal to the number of galaxies down to a limiting
magnitude of $-21^m$. We determine the error of $L_K$ by subtracting
the luminosity of each galaxy (one after another) from the total luminosity of the
cluster and averaging the resulting deviations. In the Table 2 $f_E$ denotes
the fraction of early-type galaxies down to a limiting magnitude of $M^*_K+1$
and the $N_E$ is the number of such galaxies and is given in the parentheses.

\section{CONCLUSIONS}

The Ursa Major supercluster is a layered system consisting of
large filamentary structures 
\linebreak \mbox{(see [5, Figs. 1a-1c])}
with the mean redshifts of  0.051, 0.061, and 0.071.
Only three clusters in this system, A1377, A1436, and Anon1, are
sufficiently rich both in terms of the number of galaxies inside the virial
radius and in terms of X-ray luminosity (Table~\ref{data1:Kopylova_n}).
UMa is nevertheless a supercluster of special interest, because it is a rather
isolated system 
\mbox{(see [5, Figs. 1a-1c])}
with no rich X-ray
clusters of galaxies or other superclusters in its vicinity, i.e., UMa is
located in the region of the Universe with a low density of clusters of
galaxies \cite{Tikhonov:Kopylova_n}. One may say that the UMa supercluster is
located in a lower-than-average density region and such a location must have
an effect on the entire system of clusters as a whole. The dynamic evolution
in the regions of high density of galaxies, such as superclusters, is believed
to begin very early and continue until the present time (see,
e.g., \cite{Einasto2:Kopylova_n}). In this paper we study and compare some of
the properties of virialized regions of clusters of galaxies located in the
vicinity of the UMa supercluster, both in its central part with a
factor-of-three galaxy overdensity, and in its periphery with radius of
75~Mpc. In this study we use the data of the SDSS and 2MASS catalogs. Below we
compare our results for two samples of clusters.

(1) The Schechter functions for the composite luminosity functions of
virialized regions of clusters of galaxies have similar parameters
($M^*_K$,$\alpha$) in UMa and in its neighborhood, and $M^*_K$
for the composite LF is $0.15^m$ lower than the corresponding parameter for the
LF of field clusters \cite{Kochanek:Kopylova_n}.

(2) The Schechter functions for the composite luminosity functions of
early- and late-type galaxies in virialized regions of clusters also have
similar  ($M^*_K$, $\alpha$) parameters in UMa and in its vicinity, and their
$M^*_K$ are lower than the corresponding parameter for field galaxies
\cite{Kochanek:Kopylova_n}. The fraction of early-type galaxies is higher by
$40\%$ in UMa than in clusters located in its vicinity. The number of spiral
galaxies in UMa is greater by $15\%$, and the number of such galaxies
increases ($\alpha>1$), whereas that of early-type galaxies decreases
($\alpha<1$) with increasing galaxy magnitude.

(3) The fraction of early-type galaxies down to a limiting magnitude of
$M^*_K+1$, which is equal to  $-23\fm29,$ is of about $70\%$ in UMa clusters and
in the clusters located in the UMa neighborhood. This fraction does not increase
with increasing cluster mass in the cluster mass interval
$10^{13}$~$M_{\odot}$ < M < $5.0 \times10^{14}$~$M_{\odot}$. The
average parameters of early-type galaxies are almost the same for
clusters located in the central dense region of UMa and in its nearest
vicinity, as well as in poor and rich clusters.

(4) The main parameters of clusters inside $R_{200}$ in the UMa
supercluster region  ($L_{K,200},$ $M_{200}/L_{K,200},$ and $N_{200}$) increase
with the mass of the system. The forms of the relations between these parameters
agree within the errors with the relations obtained for other samples of
clusters of galaxies  drawn with no regard to the possible membership in
superclusters. The clusters located in the densest region and identified as the
UMa supercluster exhibit twice smaller scatter in all relations compared to the
corresponding scatter for field clusters located within 75~Mpc of the
supercluster.

\begin{acknowledgments}
This work was supported by the Russian Foundation for Basic Research (grant
no.~07-02-01417a). \\
This research has made use the NASA/IPAC Extragalactic
Database (NED, {\tt http://nedwww.ipac.caltech.edu/}), which is
operated by the Jet Propulsion Laboratory, California Institute of Technology,
under contract with the NASA; Sloan Digital Sky Survey (SDSS, \linebreak {\tt
http://www.sdss.org/}); data products from the Two Micron All Sky Survey
(2MASS, {\tt  http://www.ipac.caltech.edu/2mass/releases/allsky/},
which is a joint project of the University of
Massachusetts and Infrared Processing and Analysis Center/Caltech, funded by
NASA and NSF, and X-Rays Clusters Database (BAX, {\tt
http://bax.ast.obs-mip.fr/}), which is operated by the Laboratoire
d'Astrophysique de Tarbes-Tou\-louse (LATT), under contract with the Centre
National d'Etudes Spatiales (CNES).
Funding for the creation and distribution of the SDSS Archive has been
provided by the Alfred P. Sloan Foundation, the Participating Institutions,
the National Aeronautics and Space Administration, the National Science
Foundation, the US Department of Energy, the Japanese Monbukagakusho, and
the Max Planck Society.
\end{acknowledgments}

\newpage
\begin{center}
\refname
\end{center}

\end{document}